
\input harvmac
\Title{\vbox{\hbox{HUTP-95/A017, LBL-37271, UCB-PTH-95/16}
\hbox{\tt hep-th/9505183}}}
{All Loop $N=2$ String Amplitudes}
\bigskip
\centerline{Hirosi Ooguri\foot{On leave of absence
from {\it Research Institute for Mathematical Sciences,
Kyoto University}.}}
\bigskip\centerline{\it Department of Physics, University of
California at Berkeley}
\centerline{\it 366 Le\thinspace Conte Hall, Berkeley, CA 94720}
\centerline{and}
\centerline{\it Theoretical Physics Group, Mail Stop 50A-3115}
\centerline{\it Lawrence Berkeley Laboratory,
    Berkeley, California 94720}

\vskip .1in
\centerline{and}

\vskip .1in

\centerline{Cumrun Vafa}
\bigskip\centerline{\it Lyman Laboratory of Physics}
\centerline{\it Harvard University}\centerline{\it Cambridge, MA 02138}

\vskip .3in
Using the $N=4$ topological reformulation of $N=2$ strings,
we compute all loop partition function for special compactifications
of $N=2$ strings
as a function of target moduli.  We also reinterpret $N=4$
topological amplitudes in terms of slightly modified
$N=2$ topological amplitudes.  We present some preliminary
evidence for the conjecture that $N=2$ strings is the large
$N$ limit of Holomorphic Yang-Mills in 4 dimensions.

\Date{\it {May 1995}}

\vskip 1in

\centerline{\bf Disclaimer}

\vskip .2in

This document was prepared as an account of work sponsored by the United
States Government. While this document is believed to contain correct
 information, neither the United States Government nor any agency
thereof, nor The Regents of the University of California, nor any of their
employees, makes any warranty, express or implied, or assumes any legal
liability or responsibility for the accuracy, completeness, or usefulness
of any information, apparatus, product, or process disclosed, or represents
that its use would not infringe privately owned rights.  Reference herein
to any specific commercial products process, or service by its trade name,
trademark, manufacturer, or otherwise, does not necessarily constitute or
imply its endorsement, recommendation, or favoring by the United States
Government or any agency thereof, or The Regents of the University of
California.  The views and opinions of authors expressed herein do not
necessarily state or reflect those of the United States Government or any
agency thereof, or The Regents of the University of California.

\vskip 2in

\centerline{\it Lawrence
Berkeley Laboratory is an equal opportunity employer.}

\vfill

\newsec{Introduction}
One of the simplest types of string theories is $N=2$ string.
It lives in  four dimensions, and it has finite
number of particles in the spectrum.  Moreover it describes
self-dual geometries and Yang-Mills fields \ref\oovone{
H. Ooguri and C. Vafa, ``Geometry of N=2 Strings,'' Nucl. Phys.
B361 (1991) 469} \ref\oovtwo{H. Ooguri and C. Vafa,
``N=2 Heterotic Strings,'' Nucl.
Phys. B367 (1991) 83.}, which
are conjectured to describe, through reduction, all 2 and 3 dimensional
integrable models.  Moreover the 4 dimensional $N=2$ string
itself seems to correspond to an integrable theory,
as is evidenced by perturbative vanishing of scattering amplitudes
beyond three point functions.

Given all these connections, it seems very important to understand
$N=2$ string amplitudes.  In this paper we consider this question
and find, rather surprisingly, that one can compute, at least
in special cases, the all genus partition function of $N=2$ strings.
This seems to be another evidence for the quantum integrability
of self-dual theories.  More specifically we consider compactifications
of $N=2$ strings on $T^2\times R^2$.  Using the reformulation
of $N=2$ strings in terms of $N=4$ topological strings \ref\BV{
N. Berkovits and C. Vafa, ``N=4 Topological Strings,''
Nucl. Phys. B433 (1995) 123.}, allows
one to develop techniques to compute it.

For low genus, this can be done more or less directly, because
the structure of the amplitudes are so simple.  However for $g \geq 3$
the story gets more complicated.  In such cases we have found
a modified version of the harmonicity equation of \BV\ for
which the boundary contributions cancel, and are strong enough
to yield the genus $g$ partition function up to an overall constant.
Specialized to $g=1,2$ this result agrees with explicit computations
of the amplitudes.  This is somewhat analogous to the method
used in \ref\holan{M. Bershadsky, S. Cecotti, H. Ooguri
and C. Vafa,``Holomorphic Anomalies in Topological
Field Theories,'' Nucl. Phys. B405 (1993) 279\semi
``Kodaira-Spencer Theory of Gravity and Exact Results for Quantum
String Amplitudes,'' Commun. Math. Phys. 165 (1994) 311.}\ to compute the
topological $N=2$ string amplitudes, with the replacement of
holomorphic anomaly with harmonicity equation.

Another aspect of $N=2$ string, is the topological interpretation
of what it is computing.  We show that quite generally $N=4$ topological
strings, are a slightly (but crucially) modified form of $N=2$ topological
string amplitudes.  This allows us to give a more clear interpretation
of what topological quantities the partition function computes.
In particular we see quite explicitly in the cases of genus 1 and 2
in the example of $T^2\times R^2$ what these topological quantities are,
and moreover reproduce in yet another way,
the partition function itself by direct topological evaluation.

Given that $N=2$ string has finite number of particles
it is a candidate for a search for a large N limit of a gauge
theory.  We look for this and find some preliminary
evidence that the large $N$ limit of
Holomorphic Yang-Mills theory in 4 dimensions
(2 complex dimensions)\ref\park{J.-S. Park,
``Holomorphic Yang-Mills Theory on Compact K\"ahler Manifolds,''
Nucl. Phys. B423 (1994) 559\semi S. Hyun and J.-S. Park,``
Holomorphic Yang-Mills Theory and Variation of the Donaldson Invariants,''
hep-th 9503036.}\ref\witun{E. Witten, Unpublished.}\
is $N=2$ strings.  This theory is a deformation of $N=2$ topological
Yang-Mills theory, which has been recently solved in the work of
Seiberg and Witten \ref\SeW{N. Seiberg and E. Witten, ``Electric-magnetic
Duality,
Monopole Condensation, and Confinement in N=2 Supersymmetric Yang-Mills
Theory,'' Nucl. Phys. B426 (1994) 19.}.
  This is another exciting link with $N=2$ strings.

The organization of this paper is as follows:
In section 2 we review relevant aspects of $N=2$ strings as well
as its topological reformulation.  We also give a connection between $N=4$
and $N=2$ topological amplitudes in this section.
In section 3 we show how the modified harmonicity equation
manages to avoid boundary contributions (with some of the details
postponed to appendix A).  In section 4 we consider the target
space to be $T^2\times R^2$ and evaluate the partition function
for all $g$.  We do genus 1 and 2 explicitly (with some
of the details for the genus 2 case postponed to the appendix B)
and then use
the harmonicity equation to rederive these results as well
as generalize to all $g$.  In section 5, using the topological
reinterpretation, we compute the genus 1 and genus 2 contributions
topologically and find agreement with the computation of the previous section.
Finally in section 6 we present our conclusions and conjectures.

\newsec{Review of $N=2$ Strings}
In this section we briefly review aspects of $N=2$ strings which are
relevant for this paper.  $N=2$ string was first studied in the
early days of string theory \ref\adet{M. Ademollo, L. Brink, A. D'Adda,
R.D'Auria, E. Napolitano, S. Sciuto, E. Del Guidice, P. Di Vecchia,
S. Ferrara, F. Gliozzi,, R. Musto and R. Pettorino,
``Supersymmetric String and Colour Confinement,'' Phys. Lett.
62B (1976) 105;
\semi
M. Ademollo, L. Brink, A. D'Adda,
R.D'Auria, E. Napolitano, S. Sciuto, E. Del Guidice, P. Di Vecchia,
S. Ferrara, F. Gliozzi,, R. Musto and R. Pettorino, J.H. Schwarz,
``Dual String with $U(1)$ Colour Symmetry,'' Nucl. Phys.
B111 (1976)77
\semi D.J. Bruth, D.B. Fairlie and R.G. Yates,
``Dual Models with a Colour Symmetry,''  Nucl. Phys. B108 (1976) 310.}\
and
its study was resumed with the surge of interest in string theory
\ref\liz{A.D'Adda and F. Lizzi,
``Space Dimensions from Supersymmetry for the $N=2$
Spinning String: A Four Dimensional Model,''
Phys. Lett. B191 (1987) 85\semi
M. Corvi, V. A. Kostelecky and P. Moxhay,
``Toy Superstrings,'' Phys. Rev. D39 (1989) 1611
\semi
S.D. Mathur and S. Mukhi,
``The $N=2$ Fermionic String,'' Nucl. Phys. B302 (1988) 130\semi
N. Ohta and S. Osabe,
``Hidden Extended Superconformal Symmetries in Superstrings,''
Phys. Rev. D39 (1989) 1641.}.  It was discovered
relatively
recently \oovone ,\oovtwo\
 that $N=2$ string theory has a rich geometric structure
related to self-duality phenomena.  In particular its critical dimension
is four (2 complex dimensions), and the closed string theory describes
self-dual gravity, whereas heterotic and open string versions
describe self-dual gauge theories in four dimensions
coupled to self-dual gravity.  Some of these
aspects were further studied \ref\oth{N. Marcus,``The N=2 Open String''
Nucl. Phys. B387 (1992) 263\semi
W. Siegel,``N=2, N=4 String Theory is Selfdual N=4 Yang-Mills Theory,'' Phys.
Rev. D46 (1992) 3235.}.  More recently it was shown \BV\
that the loop amplitude computations in $N=2$ theories
can be simplified by proving their equivalence to a new topological
string based on the small $N=4$ superconformal algebra.
In this way the ghosts are eliminated and at the same time
the matter fields are topologically twisted; this makes
computations much easier.  The main aim of this paper is
to further elaborate on the meaning of the $N=2$ string amplitudes
in light of this development.  In this section we will give
a brief review of the topological reformulation of \BV\
referring the interested readers for the detail to that paper.
We will mainly concentrate on the closed string case.  The
generalization to other cases (heterotic and open) are straight
forward.

$N=2$ strings are obtained by gauging the $N=2$ local supersymmetry
on the worldsheet.  This consists of the metric $g_{\mu \nu}$, two
supersymmetric partners of spin $3/2$, $\psi^{\pm}_{\mu \alpha}$
and one $U(1)$ gauge field $A_{\mu}$.  In the standard fashion,
these give rise to a pair of fermionic ghost $(b,c)$ of spin 2,
two pairs of bosonic superghosts ($\beta^{\pm},\gamma^{\pm}$)
of spin 3/2 and another pair of fermionic ghost $(\tilde b ,\tilde c)$
of spin 1.  The total ghost anomaly is $c=-6$, which is cancelled
by a matter with $c=6$, corresponding to a superconformal
theory in 4 dimensions.
The vacua of $N=2$ strings consist of theories in 4d which
have Ricci-flat metric \oovone .  These theories will necessarily
have an extended symmetry, by including the spectral flow
operators, to the small $N=4 $ superconformal algebra with
$c=6$ ($\hat c =2$).

The $N=4$ algebra consists of an energy momentum tensor
$T$ of spin 2, an $SU(2)$ current algebra of spin 1, whose
generators are denoted by $J^{++},J,J^{--}$ and 4 spin 3/2
supercurrents which form two doublets $(G^-,{\tilde G^+})$
and $({\tilde G^-, G^+})$ under the $SU(2)$ currents.
The supercurrents within a doublet have no singularities
with each other, while the oppositely charged supercurrents
of the different doublets have singular OPE (and in particular
give the energy momentum tensor).  Moreover $G^+$ and ${\tilde G^+}$
have a singular OPE with a total derivative as the residue:
$$G^+(z) {\tilde G}^+(0)\sim {\partial J^{++}(0)\over z}$$
Note in addition that
\eqn\simp{\tilde G^+ = G^-(J^{++})}
which follows
from the fact that $(G^-,{\tilde G^+})$ form an $SU(2)$ doublet.
Also note that $J^{++}$ is the left-moving spectral flow operator.
This in particular implies that the chiral field $V$ corresponding
to the volume form of the superconformal theory can be written as
\eqn\vol{V=J^{++}_L J^{++}_R}
Together with \simp\ this means that
\eqn\soi{G^-_L G^-_R V(z,\bar z)=\tilde G^+_L \tilde G^+_R(z,\bar z)}
It is important to note that the choice of two doublets
among the four supersymmetry currents is ambiguous:
In particular there is a sphere worth of inequivalent choices
given by
$$\widehat{{\tilde G}^+}(u)=u^1 {\tilde G}^+ +u^2 { G^+}$$
$$\widehat {G^-}(u)=u^1 { G}^--u^2{\tilde G^-}$$
$$\widehat {{\tilde G}^-}(u)=u^{2*}{\tilde G}^- - u^{1*}{ G^-}$$
\eqn\rot{\widehat {G^+}(u) =u^{2*}{G}^+ +u^{1*} {\tilde G^+}}
where
$$|u^1|^2+|u^2|^2=1$$
and where the complex conjugate of $u_a$ is $\epsilon^{ab} u^*_b$
(i.e. $\overline {(u^1)}=u^{2*}$ and $\overline{(u^2)}=-u^{1*}$
where $*^2=-1$).  Note that we could do this rotation for
left and right $N=4$ algebras independently, and we will use
$u_L,u_R$ to denote the left- and right-moving choices for the
rotation.

A theory with $N=4$ superconformal structure can be deformed,
preserving the $N=4$ structure using chiral field of (left,right) charge
(1,1).  There are four deformations
that can be made out of a given chiral field $\phi^i$:
$$S\rightarrow S+\int t_i^{11}G^-_LG^-_R \phi^i-t_i^{21}{\tilde G}^-_LG^-_R
\phi^i-t_i^{12}G^-_L{\tilde G}^-_R
\phi^i+t_i^{22}{\tilde G}^-_L{\tilde G}^-_R
\phi_i$$
Note that for unitary $N=4$ theories, these deformations are pairwise
complex conjugate.  In particular there exists a matrix $M_i^{j*}$
so that
$$t_i^{ab}=\epsilon^{ac} \epsilon^{bd}M_i^{j*}\bar{t_j^{cd}}$$
with $MM^*=1$.

The $N=2$ string amplitudes are computed by integration
of the string measure over the $N=2$ supermoduli.  The bosonic
piece of this moduli consists of the moduli of genus $g$ Riemann
surfaces as well as the $g$-dimensional moduli of $U(1)$ bundles with a given
instanton number $n$. For a fixed instanton number the
dimension of $\beta^{\pm}$ zero modes gives the dimension
of supermoduli.  Since they are charged under the $U(1)$ this
dimension will depend on the instanton number.  In particular
the dimension of these supermoduli is $(2g-2-n,2g-2+n)$ for
the $(\beta^+,\beta^-)$ zero modes.  In particular this means
that $|n|\leq 2g-2$ in order to get a non-zero measure.
Even though geometrically not obvious, it turns out that
we can also assign independent left-moving and right-moving instanton numbers.
So at each genus $g$ we have to compute the string amplitudes $F^g_{n_L,n_R}$
with $-2g+2\leq n_L,n_R \leq 2g-2$.  It is convenient to collect
these amplitudes in terms of a function on $u$--space.  Let
$$
\eqalign{ &F^g(u_L,u_R)=\cr
&= \sum_{-2g+2\leq n_L,n_R \leq 2g-2}{4g-4 \choose
2g-2+n_L}{4g-4\choose 2g-2+n_R} \cdot
F^g_{n_L,n_R} \times \cr
&~~~~~~~~~\times(u^1_L)^{2g-2+n_L} (u^1_R)^{2g-2+n_R}
(u^2_L)^{2g-2-n_L}(u^2_R)^{2g-2-n_R}\cr}$$

The result of \BV\ is that $F^g$ can be computed by
\eqn\nfour{
\eqalign{ F^g(u_L,u_R) = &
\int_{{\cal M}_g}
 \langle [ \prod_{A=1}^{3g-3}(\mu_A, \widehat{G_L^-}(u_L))
(\bar{\mu}_A, \widehat{G_R^-}(u_R))]
 \int_\Sigma J_L J_R \times \cr
&~~~~~\times \big[ \int_{\Sigma} \widehat{\tilde{G}_L^+}(u_L)
\widehat{\tilde{G}_R^+}(u_R) \big]^{g-1} \rangle \cr}}
where $\Sigma$ denotes the Riemann surface and
${\cal M}_g$ denotes the moduli of genus $g$ surfaces
and $\mu_A$ denote the Beltrami differentials.   In this
expression there are no ghosts left over and the $N=4$ matter
field is topologically twisted, i.e. the spin of the fields
are shifted by half their charge, so in particular $G^+,\tilde G^+$
have spin 1 and $G^-, \tilde G^-$ have spin 2 and $J^{++}$
has spin 0 and $J^{--}$ has spin 2.

Let us give a rough outline of how the above correspondence
between $N=2$ string amplitudes and the $N=4$ topological
amplitude, defined above, arises.
The simplest case of constructing this measure corresponds
to the $n_L=n_R=2g-2$.  In this case we have no $\beta^+$ zero
modes, and $(4g-4)$ $\beta^-$ zero modes.  If we had instanton
number $(g-1)$, it would have been equivalent to twisting
the fields, by the definition of topological twisting (identifying
gauge connection with half the spin connection).  So for instanton
number $(2g-2)$, we can view the amplitudes as being computed
in the topologically twisted version
but with an addition of $(g-1)$ instanton number changing operators
inserted. Note that the matter part of the instanton number changing
operator is $J^{++}$.  In the topologically twisted measure
the ($\beta^- ,\gamma^-$) ghost system have the same spin
as $(b,c)$ and the $(\beta^+,\gamma^+)$ have the same spin as
$(\tilde b,\tilde c)$, and since they are of the opposite
statistics they cancel each other out as far
as the non-zero modes are concerned. The zero modes
can also be canceled out by a judicious choice of
the position of picture changing operators.
We have $(4g-4)$ picture changing operators inserted for integration
over the supermoduli which are accompanied from the matter sector
with $G^-$.  $(3g-3)$ of them get folded with the Beltrami differentials
in cancelling the zero modes of $b$.  The integration
over the $U(1)$ moduli is traded with integration over
$g$ operators on Riemann surfaces:  $(g-1)$ of them come from operators
where $(g-1)$ of the instanton changing operators have converted
$G^-$ into $\tilde G^+$ and the last one is simply the current $J$.
This would give the correspondence
at the highest instanton numbers and the rest are obtained
by performing an $SU(2)$ rotation on the $N=2$ string side
and seeing that it corresponds to changing the instanton numbers.

\medskip

\noindent
{\it Topological Meaning of $N=2$ String Amplitudes}

\medskip

Given the fact that the physical $N=2$ string amplitudes have been reformulated
in terms of topologically twisted $N=4$ theories, it is natural
to ask if there is any topological meaning to the latter.  Recall that if we
have any $N=2$ superconformal theory we can consider the twisted version
and couple it to topological gravity, which has critical dimension $3$.
The geometrically interesting examples of such theories are sigma models
on Calabi-Yau manifolds and depending on how the left- and right-moving
degrees of freedom are twisted we get a topological theory
which counts holomorphic maps (A-model) or quantizes the variations
of complex structure on the Calabi-Yau (the Kodaira-Spencer theory
\holan\
obtained from B-model).   If the complex
dimension of Calabi-Yau is not equal to three
the topological string amplitude vanishes because the $(3g-3)$ negative
charges of the $G^-$ insertions is not balanced by the $d(g-1)$
charge violation of the $U(1)$ of the $N=2$ algebra if $d\not= 3$.
Only in the case of complex dimension 2 one
can still try to get a non-vanishing amplitude by inserting $(g-1)$ chiral
operators to the action of the form $G^-_LG^-_R V$ where $V$ is the
unique chiral
field with charge two\foot{In dimension bigger than 3 we need a negative
charged chiral field which does not exists, and in dimension 1, there
is no chiral field with charge bigger than one.} and it corresponds
to the volume form of the complex 2d manifold.  Note using
\soi\ that these $(g-1)$ insertions are the same as the $(g-1)$
insertions of $\tilde G^+_L \tilde G^+_R$.  In other words it
gives exactly the same result as the
partition function for the highest instanton number of the
$N=2$ string \nfour\ with the exception of the insertion
of $\int J_L J_R$.  It was argued in \holan\ that
this $N=2$ topological amplitude vanishes even
with this charge insertion.  In fact it was directly argued in
\BV\ that this follows rather simply from the underlying
$N=4$ algebra.  So the $N=2$ string amplitude manages
to be non-trivial precisely because of the extra insertion
of $\int J_L J_R$.  Therefore there must be a simple
topological meaning for the highest instanton amplitude
of the $N=2$ string.

For concreteness let us consider the A-model version which
is set up to count the holomorphic maps from Riemann surfaces
to Calabi-Yau manifolds.  In the limit that $\bar t_i\rightarrow \infty$
one can show that the measure is concentrated near the holomorphic
maps \holan\ .  In this case we are considering holomorphic maps
which map the Riemann surface with $(g-1)$ points on the Riemann surface
mapped to specific $(g-1)$ points on the target which is dual
to the volume form.  Actually to go to the Poincare dual of the volume form
one has to use $G^+$ trivial operators to deform the field, but that may
change the amplitude in this case because we have $J$ insertion
which does not commute with $G^+$. So we have to use the precise
representative given by $G^-_L G^-_R V$.  Each time we choose
a cohomology representative in target of degree $d$ (corresponding
to $d$-forms), it gives rise to a $d-2$ form on moduli space
(which translating  degree to charge,
in the operator language means that the charge is decreased by
two units because of the insertion of $\oint G^-_L
\oint G^-_R$)\foot{The form on the moduli space
can be described by considering
the canonical map from
the total space of the Riemann surface and the moduli space of holomorphic
maps to the target manifold, and using the pull-back of the $d$-form and
integrating it over the Riemann surface.}.
In our case each volume form will give a $(1,1)$ form
on the moduli space of holomorphic maps which we denote
by $k$.  So consider the moduli space
${\cal M}^g$
of holomorphic
maps from genus $g$ to the 2 complex manifold.
  The formal complex dimension of ${\cal M}$ is $(g-1)$,
however it typically has a dimension bigger than $(g-1)$.  In such
cases the topological amplitude computation is done by considering
the bundle ${\cal V}$ on ${\cal M}$ whose fibers are the anti-ghost zero
modes which is $H^1(N)$ where $N$ is the pull back
of the normal bundle piece of the tangent bundle
on the manifold restricted to the
holomorphic image of the Riemann surface.  Let $n$ be
the dimension of ${\cal V}$.  Then the complex dimension of ${\cal M}$
is $(g-1+n)$.
Therefore if it were not for the $\int J_LJ_R$ insertion,
the usual arguments of topological strings, in the simple cases,
would lead to the computation of
\eqn\simc{
 \int_{{\cal M}}k^{g-1}c_n({\cal V})}
where $c_n$ denotes the $n$-th chern class of ${\cal V}$.
However as mentioned before this amplitude
vanishes.  The effect of the $\int J_LJ_R$ insertion,
will correspond on the moduli of holomorphic maps to a (1,1) form which
we denote by ${\cal J}$. This has the effect of absorbing one of
the fermion zero modes which was responsible for the vanishing of the
amplitude.  Thus the characteristic class that we will end up with
from ${\cal V}$
will be of dimension $(n-1)$.  The precise form of it may depend
on the case under consideration.
 Therefore using the same reasoning as for
topological theories we see that the top instanton number amplitude
for $N=2$ strings in the ${\bar t}\rightarrow \infty$ computes
\eqn\topeq{F^g_{2g-2,2g-2}\big|_{\bar t \rightarrow \infty}=
 \int_{{\cal M}^g}
k^{g-1}\wedge c_{n-1}({\cal V})\wedge
{\cal J}}
Later in this paper we will see how this works in detail in the case
of the four manifold $T^2\times R^2$ for $g=1,2$.
It happens that for some topological strings the formula \simc\ is
modified.  An example of this is discussed in \ref\wiwzw{E. Witten,
``The N Matrix Model and Gauged WZW Models,'' Nucl. Phys. B371 (1992) 191.}.
  In such cases some of the insertions
of operators corresponding to fields (the analog
of $k$ in the above) will be replaced by modifying the bundle {\cal V}.
It turns out that this does happen for us for $g\geq 3$ for the example
of $T^2\times R^2$.  For $g \geq 3$ the above formula in this
case gets replaced by
\eqn\topp{F^g_{2g-2,2g-2}\big|_{\bar t \rightarrow \infty}=
 \int_{{\cal M}^g}
k\wedge c_{n+g-3}({\widetilde {\cal V}})\wedge
{\cal J}}
for some $\widetilde {\cal V}$.  Unfortunately there is no
general prescription for computing this that we are aware of,
and it very much depends on the models.  We have not
computed $\widetilde {\cal V}$ for $T^2\times R^2$, which is
relevant for $g>2$ amplitudes.

\newsec{Harmonicity Equation}

In this section, we will prove that the $g$--loop amplitude
$F^{g}(u_L,u_R)$ solves the equations
\eqn\harmonicL{
\epsilon^{ab} u_R^c  {\partial \over \partial u_L^a}
                 D_{t^{bc}} F^{g} (u_L,u_R) = 0 }
\eqn\harmonicR{
 \epsilon^{ab} u_L^c  {\partial \over \partial u_R^a}
                 D_{t^{cb}} F^{g} (u_L,u_R) = 0. }
In the paper \BV, Berkovits
and one of the authors have pointed out that the stronger version
of these equations
\eqn\strongharmonic{
\eqalign{ & \epsilon^{ab} {\partial \over \partial u_L^a}
                 D_{t^{bc}} F^{g} (u_L,u_R) = 0 \cr
          & \epsilon^{ab} {\partial \over \partial u_R^a}
                 D_{t^{cb}} F^{g} (u_L,u_R) = 0 \cr} }
would hold if the contributions from the
boundary of the moduli space ${\cal M}_g$ and contact
terms in operator products were absent. The purpose of this section
is to examine these contributions carefully. As we shall see,
there are in fact contact terms which spoil \strongharmonic.
This is also in accord with the fact that if there were no corrections
to \strongharmonic\ it would lead to puzzling conclusions
\ref\nberk{N. Berkovits,``Vanishing Theorems for the Selfdual N=2 String'',
hep-th/9412179.}.
Fortunately these contact terms are cancelled out in
\harmonicL\ and \harmonicR. In the following, we will
refer to these as {\it harmonicity equations}.

In the next section,
we will examine the case when the target space is
$M= T^2 \times R^2$. For $g=2$, we can compute
$F^{g}$ directly and check explicitly that the
$F^{g}$ directly and check explicitly that the
harmonicity equations \harmonicL\
and \harmonicR\ are satisfied. Furthermore
the harmonicity equations will make it possible
to determine $F^{g}$ for all $g \geq 3$ up to a
constant factor independent of target space moduli
at each $g$.

Now let us prove the harmonicity equations. The steps
in the proof are parallel to those used in \BV\ except
we will have to be very careful with many boundary contributions
that arise.
 The covariant
derivative $D_{t^{ab}}$ is defined so that its action on
$F^{g}$ generates an insertion of a marginal operator
corresponding to the target space moduli $t^{ab}$;
$$\eqalign{& w_L^a w_R^b D_{t^{ab}} F^{g}(u_L,u_R) = \cr
&= \int_{{\cal M}_g}
 \langle [ \prod_{A=1}^{3g-3}(\mu_A, \widehat{G_L^-}(u_L))
(\bar{\mu}_A, \widehat{G_R^-}(u_R))]
 \int_\Sigma J_L J_R \times \cr
&~~~~~ \times
 \left[ \int_{\Sigma} \widehat{\tilde{G}_L^+}(u_L)
\widehat{\tilde{G}_R^+}(u_R) \right]^{g-1}
 \int_{\Sigma} \{ \widehat{\tilde{Q}_L^+}(w_L) ,
[ \widehat{\tilde{Q}_R^+}(w_R) ,
 \bar{\phi} ] \} \rangle \cr}$$
where $\bar{\phi}$ is an anti-chiral primary field coupled to
the moduli $t^{ab}$. Therefore the left-hand side of \harmonicL\
can be written as
\eqn\harmonictwo{ \eqalign{
& \epsilon^{ab} u_R^c {\partial \over \partial u_L^a}
 D_{t^{bc}} F^{g} = \cr
& = \epsilon^{ab}{\partial^2 \over \partial u_L^a \partial w_L^b}
   \left[ w_L^b u_R^c D_{t^{bc}} F^{g} \right] = \cr
& = \epsilon^{ab} {\partial^2 \over \partial u_L^a \partial w_L^b}
   \int_{{\cal M}_g}
 \langle [ \prod_{A=1}^{3g-3} (\mu_A, \widehat{G_L^-}(u_L))
(\bar{\mu}_A, \widehat{G_R^-}(u_R))] \times \cr
&~~~~~~~~~~~~~~~~~~~~~ \times
   \int_\Sigma J_L J_R
 \left[ \int_{\Sigma} \widehat{\tilde{G}^+_L}(u_L)
\widehat{\tilde{G}_R^+}(u_R) \right]^{g-1} \times \cr
&~~~~~~~~~~~~~~~~~~~~~ \times
 \int_\Sigma \{ \widehat{\tilde{Q}^+_L}(w_L)
, [ \widehat{\tilde{Q}^+_R}(u_R) , \bar{\phi} ] \}
\rangle .\cr}}
Note that the marginal operator inserted here is
$\{ \widehat{\tilde{Q}^+_L}(w_L) , [
\widehat{\tilde{Q}^+_R}(u_R), \bar{\phi}] \} $
with $u_R$ in the right-mover
while $w_L \neq u_L$ in the left-mover.
Since
$$ \eqalign{ &\epsilon^{ab}
 {\partial^2 \over \partial u_L^a \partial w_L^b}
  \widehat{\tilde{G}_L^+}(u_L) \widehat{\tilde{Q}_L^+}
(w_L)
 = \tilde{G}_L^+ Q_L^+ -
   G_L^+ \tilde{Q}_L^+ \cr
&~~~~~=(\epsilon_{ab} u_L^a w_L^b)^{-1}
  \left( \widehat{\tilde{G}_L^+}(u_L)
\widehat{\tilde{Q}_L^+}(w_L)
  - \widehat{\tilde{G}_L^+}(w_L)
\widehat{\tilde{Q}_L^+}(u_L) \right) \cr}$$
and similarly
$$ \eqalign{& \epsilon^{ab}
 {\partial^2 \over \partial u_L^a \partial w_L^b}
  \widehat{G_L^-}(u_L) \widehat{\tilde{Q}_L^+}
(w_L)
 = G_L^- Q_L^+ +
   \tilde{G}_L^- \tilde{Q}_L^+ \cr
&~~~~~= (\epsilon_{ab} u_L^a w_L^b)^{-1}
  \left( \widehat{G_L^-}(u_L)
\widehat{\tilde{Q}_L^+}(w_L)
  - \widehat{G_L^-}(w_L)
\widehat{\tilde{Q}_L^+}(u_L) \right) , \cr}$$
the differential operator $\epsilon^{ab} \partial_{u_L^a}
\partial_{w_L^b}$
exchanges $u_L$ and $w_L$. In the following, we will show that
if we pick any of $(4g-4)$ $u_L$'s in the correlation function
$$ \eqalign{ &f^{g}(u_L,u_R;w_L)=
\int_{{\cal M}_g}
 \langle [ \prod_{A=1}^{3g-3} (\mu_A, \widehat{G_L^-}(u_L))
(\bar{\mu}_A, \widehat{G_R^-}(u_R))]
\int_\Sigma J_L J_R \times \cr
&~~~~~~~~~~~~~~~\times
 \left[ \int_{\Sigma} \widehat{\tilde{G}^+_L}(u_L)
\widehat{\tilde{G}_R^+}(u_R) \right]^{g-1}
 \int_\Sigma \{ \widehat{\tilde{Q}^+_L}(w_L)
, [ \widehat{\tilde{Q}^+_R}(u_R) , \bar{\phi} ] \}
\rangle ,\cr} $$
and exchange it with $w_L$, $f^{g}$ remains invariant.
This implies that $f^{g}$ obeys
$$  \epsilon^{ab} {\partial^2 \over \partial u_L^a \partial
w_L^b} f^{g}(u_L,u_R;w_L) = 0, $$
which is equivalent to the harmonicity equation \harmonicL\
by \harmonictwo.

Since
\eqn\commutator{
\eqalign{
& \widehat{G_L^-}(u_L) = [ \widehat{\tilde{Q}_L^+}(u_L),
 J_L^{--} ] \cr
& \widehat{\tilde{G}_L^+}(u_L)
= - [ \widehat{\tilde{Q}_L^+}(u_L),
 J_L ], \cr} }
exchanging $u_L$ and $w_L$ is same as
exchanging locations of $\widehat{\tilde{Q}_L^+}(u_L)$
and $\widehat{\tilde{Q}_L^+}(w_L)$ in
$f^{g}(u_L, u_R; w_L) $.
We can exchange their locations just like
two automobile drivers would do when they try
to pass each other on a narrow country road.
We can first move $\widehat{\tilde{Q}_L^+}(w_L)$ off
from $[\widehat{\tilde{Q}_R^+}(u_R), \bar{\phi}]$,
park it in a ``turn-out'' at $\int_\Sigma J_L J_R$. We then
move $\widehat{\tilde{Q}_L^+}(u_L)$ off from $J_L^{--}$ or
$J_L$ in \commutator,
let it pass $\widehat{\tilde{Q}_L^+}(w_L)$, and stop it at
$[\widehat{\tilde{Q}_R^+}(u_R), \bar{\phi}]$. Finally
we move $\widehat{\tilde{Q}_L^+}(w_L)$ out from the
turn-out and stop it at $J_L^{--}$ or $J_L$. In this way,
we can exchange locations of
$\widehat{\tilde{Q}_L^+}(w_L)$
and $\widehat{\tilde{Q}_L^+}(u_L)$.

This is not the complete story since
we have neglected the anti-commutators
of $\widehat{\tilde{Q}_L^+}$ with
$\widehat{G_L^-}$ and $\widehat{\tilde{G}_L^+}$;
$$\eqalign{ &
  \{ \widehat{\tilde{Q}_L^+}(w_L) ,
     \widehat{G_L^-}(u_L) \} =
   2 (\epsilon_{ab} u_L^a w_L^b) T_L \cr
&
  \{ \widehat{\tilde{Q}_L^+}(w_L) ,
     \widehat{\tilde{G}_L^+}(u_L) \} =
   {1 \over 2} (\epsilon_{ab} u_L^a w_L^b) \partial J_L^{++}, \cr}
$$
which appear
when we move $\widehat{\tilde{Q}_L^+}(u_L)$ and
$\widehat{\tilde{Q}_L^+}(w_L)$  back and forth in
$f^{g}$. Therefore what we have shown so far is
that \harmonictwo\ is a linear
combination of the following four types of terms;
\eqn\first{
\eqalign{ &
\int_{{\cal M}_g}
 \langle (\mu_{A'}, T_L)
 \prod_{A\neq A'} (\mu_A, \widehat{G_L^-}(u_L)) \prod_{A=1}^{3g-3}
(\bar{\mu}_A, \widehat{G_R^-}(u_R)) \times \cr
&~~~~~~~~\times
 \int_\Sigma J_L J_R
 \left[ \int_{\Sigma} \widehat{\tilde{G}^+_L}(u_L)
\widehat{\tilde{G}_R^+}(u_R) \right]^{g-1}
 \int_\Sigma [ \widehat{\tilde{Q}^+_R}(u_R) , \bar{\phi} ]
\rangle ,\cr}}
\eqn\second{
\eqalign{ &
\int_{{\cal M}_g}
 \langle (\mu_{A'}, T_L)(\mu_{A''},J_L^{--})
 \prod_{A\neq A',A''} (\mu_A, \widehat{G_L^-}(u_L)) \prod_{A=1}^{3g-3}
(\bar{\mu}_A, \widehat{G_R^-}(u_R)) \times \cr
&~~~~~~~~\times
 \int_\Sigma J_L J_R
 \left[ \int_{\Sigma} \widehat{\tilde{G}^+_L}(u_L)
\widehat{\tilde{G}_R^+}(u_R) \right]^{g-1}
 \int_\Sigma \{ \widehat{\tilde{Q}_L^+}(u_L), [ \widehat{\tilde{Q}^+_R}(u_R)
, \bar{\phi} ] \}
\rangle ,\cr}}
\eqn\third{
\eqalign{ &
\int_{{\cal M}_g}
 \langle [ \prod_{A=1}^{3g-3} (\mu_A, \widehat{G_L^-}(u_L))
(\bar{\mu}_A, \widehat{G_R^-}(u_R))]
\int_\Sigma J_L J_R \cr
&~~~~~~~~ \times
\int_\Sigma \partial J_L^{++} \widehat{\tilde{G}_R^+}(u_R)
 \left[ \int_{\Sigma} \widehat{\tilde{G}^+_L}(u_L)
\widehat{\tilde{G}_R^+}(u_R) \right]^{g-2}
 \int_\Sigma [ \widehat{\tilde{Q}^+_R}(u_R) , \bar{\phi} ]
\rangle .\cr}}
\eqn\fourth{
\eqalign{ &
\int_{{\cal M}_g} \langle (\mu_{A'}, J_L^{--})
 \prod_{A\neq A'} (\mu_A, \widehat{G_L^-}(u_L))
\prod_{A=1}^{3g-3}
(\bar{\mu}_A, \widehat{G_R^-}(u_R)) \times \cr
&~~~~~~~~\times
 \int_\Sigma \partial J_L^{++} J_R
 \left[ \int_{\Sigma} \widehat{\tilde{G}^+_L}(u_L)
\widehat{\tilde{G}_R^+}(u_R) \right]^{g-1}
 \int_\Sigma  [ \widehat{\tilde{Q}^+_R}(u_R) ,
\bar{\phi} ]
\rangle ,\cr}}
where $A',A''=1,...,3g-3$ ($A'\neq A''$).
We did not write terms which are related
to one of these four types by contour deformation
of the operators. To prove
the harmonicity equation \harmonicL, we want to show
that these four terms vanish.

The first two, \first\ and \second, contain $(\mu_{A'}, T_L)$
thus are total derivative in the moduli space ${\cal M}_g$
of smooth Riemann surfaces.  Then these integrals reduce
to integrals on the boundaries of ${\cal M}_g$. In Appendix
A, it is shown that there is no boundary contribution
to these integrals and therefore we can ignore
\first\ and \second.

The third \third\ and fourth \fourth\ will also be
zero if we can integrate away the total derivatives
$\partial J_L^{++} \widehat{\tilde{G}_R^{+}}(u_R)$
and $\partial J_L^{++} J_R$. This will be
possible if there is no singularity in the domains of
these integrals.
Let us first consider the third term \third.
By the Cauchy theorem, the surface integral of
$\partial J_L^{++} \widehat{\tilde{G}_R^{+}}(u_R)$ becomes
anti-holomorphic contour integrals around other operators
in the correlation function. Although there
are operators in \third\ which have operator product
singularities with
$J_L^{++} \widehat{\tilde{G}_R^{+}}(u_R)$, none of
these singularities
survive after the contour integrals since they all have
wrong powers in the holomorphic and anti-holomorphic coordinates.
Thus \third\ vanishes by integration-by-parts.

Let us examine the last piece \fourth. Again
$J_L^{++} J_R$ has singularities with other operators in
\fourth, but the only ones that survive
the contour integrals are those at
 $\widehat{G_R^-}(u_R)$, which gives
$$
\eqalign{ &
\int_{{\cal M}_g} \langle (\mu_{A'}, J_L^{--})
(\mu_{A'''}, J^{++}_L \widehat{G_R^-}(u_R))
 \prod_{A\neq A'} (\mu_A, \widehat{G_L^-}(u_L))
\prod_{A \neq A'''}
(\bar{\mu}_A, \widehat{G_R^-}(u_R)) \times \cr
&~~~~~~~~\times
 \left[ \int_{\Sigma} \widehat{\tilde{G}^+_L}(u_L)
\widehat{\tilde{G}_R^+}(u_R) \right]^{g-1}
 \int_\Sigma [ \widehat{\tilde{Q}^+_R}(u_R) , \bar{\phi} ]
\rangle ,\cr}$$
where $A'''=1,...,3g-3$
However it is easy to see that this in fact is also zero.
To show this, we can just move $\widehat{\tilde{Q}^+_R}(u_R)$ off
from $\bar{\phi}$. Since everything in the above (anti-) commutes
with $\widehat{\tilde{Q}^+_R}(u_R)$, its contour drops off from
the Riemann surface. Therefore the last term \fourth\ also vanishes
by integration-by-parts.

We have shown that \first, \second, \third\ and \fourth\ are all zero.
Thus the harmonicity equation \harmonicL\ is proven.

We should point out that in this proof it is crucial that
$\widehat{\tilde{Q}_R^+}(u_R)$ in the marginal operator
has the same $u_R$ as the rest of the operators $\widehat{G_R^-}$
and $\widehat{\tilde{G}_R^+}$ in $F^{g}$. Otherwise
the operators $J_L^{++} \widehat{\tilde{G}_R^+}$ would have
a singularity with $[\widehat{\tilde{Q}_R^+}(u_R), \bar{\phi}]$
with a non-zero pole residue, and the last part of the proof
would not go through. This is where the stronger version
of the harmonicity equation \strongharmonic\ breaks down.

\newsec{$N=2$ String Amplitudes on $T^2 \times R^2$
to All Order in Perturbation}

In this section, we will examine the $N=2$ string
amplitudes on $T^2 \times R^2$ in detail. We consider
the $A$-model only. The corresponding amplitudes
in the $B$-model are obtained by simply replacing
the K\"ahler moduli $\sigma$ by the complex moduli $\rho$.

At genus one, the string amplitude has
been computed in our previous paper \oovone\ as
$$  F^1
= -\log \Big( \sqrt{{\rm Im}\sigma {\rm Im}\rho}
|\eta(\sigma)|^2|\eta(\rho)|^2 \Big) ,$$
where $\sigma$ and $\rho$ are K\"aher and complex moduli
of $T^2$ respectively. At genus two, we will carry out
explicit computation below and derive\foot{
This is up to an overall normalization.  To obtain
the topological normalization discussed in section 5.1 we need
to multiply the above result by ${1 \over 4 (2 \pi)^4}$.}
$$ F^2(u_L, u_R) = \sum_{(n,m) \neq (0,0)}
       \left( {u^1_L u^1_R \over n + m \sigma} +
               {u^2_L u^2_R \over n + m \bar{\sigma}} \right)^4 .$$
We will show that $F^2_{2,2}$ has a nice topological interpretation
as counting of number of holomorphic maps from genus two surfaces
to $T^2$.

We will also verify that these expressions for $F^1$ and
$F^2$ are consistent with the harmonicity equations, \harmonicL\
and \harmonicR. We will then apply these equations to
$g \geq 3$ amplitudes. It turns out that the harmonicity equations
determine $F^g$ up to an overall constant at each genus as
$$  F^g(u_L, u_R) =({\rm const}) \times
 \sum_{(n,m) \neq (0,0)} |n+m\sigma|^{2g-4}
       \left( {u^1_L u^1_R \over n + m \sigma} +
               {u^2_L u^2_R \over n + m \bar{\sigma}} \right)^{4g-4}
.$$
Topological interpretation of $g \geq 3$ amplitudes will be discussed
in section 5.

\subsec{Genus One}

At genus one, the $N=2$ string amplitude $F^1$ is
given by
$$ F^1 = {1 \over 4} \int {d \tau \over ({\rm Im}\tau)^2}
\langle \Big( \int J_L J_R \Big)^2 \rangle  $$
where $\tau$ is the modulus of the worldsheet torus.
This is defined in such a way that the derivative
$D_{t^{ab}}$ with respect to the target space moduli
$t^{ab}$ gives
$$ u_L^a u_R^b D_{t^{ab}} F^1 =
 {1 \over 2} \int {d \tau \over ({\rm Im}\tau)^2}
 \langle \int J_L J_R \int \widehat{G_L^-}(u_L)
 \widehat{G_R^-}(u_R) \phi \rangle $$
which is natural generalization of \nfour\ (
$1/2$ is due to the $Z_2$ symmetry of the torus).

When the target space is $T^2 \times R^2$,
this expression reduces to
$$ F^1 = {1 \over 2} \sum_{n,m,r,s \in Z}
\int {d^2 \tau \over ({\rm Im}\tau)^2}
\exp( -S) $$
where
$$  S = {1 \over {\rm Im} \tau {\rm Im} \rho}
     \Big( t p_L \bar{p}_R + \bar{t} \bar{p}_L p_R \Big) $$
$p_L$ and $p_R$ are string momenta on $T^2$ given by
$$
\eqalign{ &p_{L}
= (n +\rho s) - (m + \rho r) \bar{\tau} \cr
&\bar{p}_{L}
= (n + \bar{\rho} s) - (m + \bar{\rho} r) \bar{\tau} \cr
&p_{R}
= (n +\rho s ) - (m + \rho r) \tau \cr
&\bar{p}_{R}
= (n +\bar{\rho} s) - (m + \bar{\rho} r) \tau \cr}
$$
with $\rho$ being the complex moduli of $T^2$, and
$t$ is the K\"ahler modulus whose real and imaginary parts are
the volume of $T^2$ and the theta parameter of the sigma-model.
In \oovone\ (See also \holan\ and
\ref\dkl{L. Dixon, V.S. Kaplunovsky and J. Louis, ``Moduli Dependence of
String Loop Corrections to Gauge Coupling Constants,''
Nucl. Phys. B320 (1989) 669.}) , this integral is evaluated with the result
\eqn\genusone{ F^1
= -\log \Big( \sqrt{{\rm Im} \sigma {\rm Im}\rho}
|\eta(\sigma)|^2|\eta(\rho)|^2 \Big) }
where $\sigma = (8\pi i)^{-1} t$.

It is easy to show that this expression is consistent
with the harmonicity equation. In fact, in this case,
the stronger version of the harmonicity equation
\strongharmonic\ also holds. On $F^1$, the stronger version
takes the form
\eqn\genusoneharmonic{ \Big( {D \over D t^{1a}}
{D \over D t^{2b'}} - {D \over D t^{2a}} {D \over D t^{1b'}}
     \Big) F^1  = 0 }
where $a,b = 1,2$.  Let us first consider the case
when $(a,b) = (1,2)$ and choose $t^{11}$ to be the K\"ahler modulus
$t$ and $t^{22'}$ to be its complex conjugate $\bar{t}$.
In this case $D_{t^{21}}$ and $D_{t^{1b'}}$ insert
operators $\partial X^{\bar{2}} \bar{\partial} X^1$
and $\partial X^2 \bar{\partial} X^{\bar{1}}$ on the worldsheet
where $X^1$ and $X^2$ are coordinates on $T^2$ and $R^2$ parts of
the target space respectively. Since there is no winding mode in
$R^2$ direction, $\partial X^{\bar{2}}$ and $\partial X^2$
are contracted according to the Wick rule, and its residue is
proportional to $\bar{\partial} X^1 \bar{\partial} X^{\bar{1}}$.
This is equal (up to a factor $(t + \bar{t})^{-1}$)
to the $T^2$ part of the
energy-momentum tensor $T_R$, which we can convert into
a derivative $\partial_{\bar{\tau}}$ with respect to the
worldsheet modulus. Thus we find
$$  D_{t^{21}} D_{t^{12'}} F^1 =
  {1 \over \pi i ( t+\bar{t} )} \sum_{n,m,r,s} \int d^2 \tau
 {\partial \over \partial \bar{\tau}} \Big(
{1 \over {\rm Im} \tau } \exp(-S) \Big) .$$
Since this integral is total derivative in
$\bar{\tau}$, it will receive contribution only from
the boundary of the moduli space at ${\rm Im} \tau \rightarrow
\infty$. There the sum over $n,m,r,s$ becomes an integral
and we obtain
$$ \sum_{n,m,r,s} \exp(-S)  \rightarrow
    { \pi {\rm Im} \tau  \over t +\bar{t}} .$$
The contribution from the boundary of the moduli space then gives
$$ D_{t^{21}} D_{t^{12'}} F^1 =
  {1 \over 2(t+\bar{t})^2 } .$$
By combining this with the harmonicity equation
\genusoneharmonic ,
$$ D_t D_{\bar{t}} F^1 = {1 \over 2 (t+\bar{t})^2} $$
and this is consistent with the expression \genusone\ of $F^1$.

By considering other cases such as $(a,b)=(1,1)$ and $t^{11}=t$,
$t^{12'} = \rho$, we can derive identities such as
$$ D_t D_\rho F^1  = 0 ,$$
which are also consistent with the expression \genusone.
In fact, the harmonicity equation \genusoneharmonic\ together
with the modular invariance in $\rho$ and the duality in
$t$ uniquely determine $F^1$ to be of the form \genusone.

\subsec{Genus Two; Evaluation I}

Genus two computation of the amplitude is much easier
than $g>2$, because all the fermionic fields in the
definition of the partition function are absorbed by the fermion
zero modes (this is mirrored, as we will explain later, in the
simplicity in its topological reinterpretation).  We leave this aspect
of the genus 2 amplitude computation to
Appendix B, where it is shown that the genus $2$ amplitude $F^{2}$
on $T^4$ is given by
\eqn\fourtorus{
\eqalign{ F^{2} =&
\sum_{P_L, P_R} \int
\left({\det g \over \det {\rm Im} \Omega} \right)^2
\langle[ \det (\widehat{P}_L+ \widehat{r}_L)
 \det (\widehat{\bar{P}}_R+ \widehat{\bar{r}}_R)]^2\rangle  \times \cr
&~~~~~~~~~~~~~~~~~~ \times
\exp[-S(P_L,P_R)] {d^3 \Omega d^3 \bar{\Omega}
\over  [ \det {\rm Im} \Omega]^3}. \cr} }
where $\widehat{P}_L$ and $\widehat{\bar{P}}_R$ are given by
$$
 \eqalign{&
 \widehat{P}_{La}^{i} = u_L^1 P_{La}^{i} + u_L^2 \epsilon^{ij} g_{j\bar{k}}
    P_{La}^{\bar{k}} \cr
  & \widehat{P}_{Ra}^{\bar{i}} =
  u_R^1 P_{Ra}^{\bar{i}} + u_R^2 \epsilon^{\bar{i}\bar{j}}
g_{\bar{j} k} P_{Ra}^{k} \cr} $$
with $P_L$ and $P_R$ being parametrized by a set of integers
$n,m,r,s$ as
$$
\eqalign{
 P_{La}^{i} & = (n^i_a + \rho^i_j s^j_a) -
                (m^{ib} + \rho^i_j r^{jb}) \bar{\Omega}_{ba} \cr
 P_{La}^{\bar{i}} & = (n^i_a + \bar{\rho}^i_j s^j_a) -
                (m^{ib} + \bar{\rho}^i_j r^{jb}) \bar{\Omega}_{ba} \cr
 P_{Ra}^{i} & =
   (n^i_a + \rho^i_j s^j_a) -
      (m^{ib} + \rho^i_j r^{jb}) \Omega_{ba} \cr
 P_{Ra}^{\bar{i}} & =
   (n^i_a + \bar{\rho}^i_j s^j_a) -
      (m^{ib} + \bar{\rho}^i_j r^{jb}) \Omega_{ba}, \cr} $$
$\widehat{r}_L$ and $\widehat{r}_R$ are quantum variables obeying
\eqn\wickrules{
 \eqalign{ & \langle \widehat{r}_{La}^{i} \widehat{r}_{Lb}^{j} \rangle
= \langle \widehat{r}_{Ra}^{\bar{i}}
    \widehat{r}_{Rb}^{\bar{j}} \rangle = 0 \cr
& \langle \widehat{r}_{La}^{i} \widehat{r}_{Rb}^{\bar{j}} \rangle
  = - g^{i \bar{j}} (u_L^1 u_R^1 + u_L^2 u_R^2)
({\rm Im} \Omega)_{ab}, \cr } }
the action $S$ is given by
$$ S(P_L, P_R)  = \Big( t_{i \bar{j}} P_{La}^i P_{Rb}^{\bar{j}}
   + \bar{t}_{i \bar{j}} P_{La}^{\bar{j}} P_{Rb}^i
  \Big) ({\rm Im} \Omega^{-1})^{ab} ,$$
with
$ t_{i \bar{j}} = g_{i\bar{j}} + i \theta^\alpha k^\alpha_{i \bar{j}} $
and $\Omega_{ab}$ is the period matrix of the genus $2$ surface.

It is straightforward to perform the Wick contraction
of $\langle[ \det (\widehat{P}_L+ \widehat{r}_L)
 \det (\widehat{\bar{P}}_R+ \widehat{\bar{r}}_R)]^2\rangle $
using \wickrules\ as
\eqn\fourtorusintegrand{
 \eqalign{
&\left( {\det g \over \det{\rm Im} \Omega} \right)^2
\langle[ \det (\widehat{P}_L+ \widehat{r}_L)
 \det (\widehat{\bar{P}}_R+ \widehat{\bar{r}}_R)]^2\rangle = \cr
&= \left( {\det g \det \widehat{P}_L \det \widehat{\bar{P}}_R
\over \det {\rm Im} \Omega} \right)^2
- \cr
&-  4
{\det g  \det \widehat{P}_L \det
\widehat{\bar{P}}_R \over \det {\rm Im} \Omega}
(\widehat{P}_L, \widehat{\bar{P}}_R)
(u_L^1 u_R^1 + u_L^2 u_R^2) + \cr
&+
\left( 16 {\det g  \det \widehat{P}_L \det
\widehat{\bar{P}}_R
\over \det {\rm Im} \Omega}
  + 2 (\widehat{P}_L, \widehat{\bar{P}}_R) \right)
(u_L^1 u_R^1 + u_L^2 u_R^2)^2 - \cr
&- 12 (\widehat{P}_L, \widehat{\bar{P}}_R)
(u_L^1 u_R^1 + u_L^2 u_R^2)^3 +
12 (u_L^1 u_R^1 + u_L^2 u_R^2)^4 \cr }}
where
$$ (\widehat{P}_L,
\widehat{\bar{P}}_R) = g_{i \bar{j}} \widehat{P}_{La}^i
({\rm Im} \Omega^{-1})^{ab} \widehat{P}_{Rb}^{\bar{j}} .$$

To compute $F^{2}$ on $T^2 \times R^2$, we set
$$ (g_{i \bar{j}}) =
 \left( \matrix{ r_1/{\rm Im}\rho
 & 0 \cr 0 & r_2 \cr} \right) $$
and send $r_2 \rightarrow \infty$ while keeping $r_1$ finite.
In order for the
action $S$ to remain finite, we must impose the momenta
in the $r_2$-direction to vanish, $P_{La}^2=P_{Ra}^2=0$.
The action then becomes
\eqn\twotorusaction{
 S = t (p_L, \bar{p}_R) + \bar{t} (\bar{p}_L , p_R) , }
where
$$ (p_L, \bar{p}_R) = {
p_{La}({\rm Im} \Omega^{-1})^{ab}
 \bar{p}_{Rb} \over  {\rm Im} \rho}, $$
$p_L$ and $p_R$ are string momenta on $T^2$ given by
$$
\eqalign{ &p_{La} = P_{La}^1
= (n_a+\rho s_a) - (m^b + \rho r^b) \bar{\Omega}_{ab} \cr
&\bar{p}_{La} = P_{La}^{\bar{1}}
= (n_a+ \bar{\rho} s_a) - (m^b + \bar{\rho} r^b) \bar{\Omega}_{ab} \cr
&p_{Ra} = P_{Ra}^1
= (n_a+\rho s_a) - (m^b + \rho r^b) \Omega_{ab} \cr
&\bar{p}_{Ra} = P_{Ra}^{\bar{1}}
= (n_a+\bar{\rho} s_a) - (m^b + \bar{\rho} r^b) \Omega_{ab} \cr}
$$
with $\rho$ being the complex moduli of $T^2$, and
$t$ is the K\"ahler moduli of $T^2$,
whose real and imaginary parts are $r_1$
and $\theta$ respectively.
In this limit, we can make the following substitutions
in \fourtorusintegrand :
\eqn\diagonal{ \eqalign{
 {\det g \det \widehat{P}_L \det \widehat{\bar{P}}_R
  \over \det {\rm Im} \Omega}
& = (t +\bar{t})^2 u_L^1 u_R^1 u_L^2  u_R^2
\Big[ (p_L, \bar{p}_R)(\bar{p}_L, p_R)-
(p_L, p_R)(\bar{p}_L, \bar{p}_R) \Big] \cr
 (\widehat{P}_L, \widehat{P}_R) &=
(t+ \bar{t}) \Big[ u_L^1 u_R^1
(p_L, \bar{p}_R) + u_L^2 u_R^2 (\bar{p}_L, p_R) \Big] .\cr} }

We can now expand $F^{2}(u_L,u_R)$ in powers in
$u_L$ and $u_R$ and extract $F_{n,m}^{2}$. Since
$F^{2}$ depends on $u_L^i$ and $u_R^i$ only
through the combinations $u_L^1 u_R^1$ and $u_L^2 u_R^2$
as one can see from \diagonal\ and \fourtorusintegrand,
the off-diagonal terms $F_{n,m}^{2}$ ($n \neq m$) all
vanish for $T^2 \times R^2$ (This is not the case
for a generic $T^4$). Since the unitarity of the sigma--model
implies $\overline{F_{n,m}^{g}} = F_{-m,-m}^{g}$, we only
need to compute $F_{2,2}^{2}$, $F_{1,1}^{2}$ and
$F_{0,0}^{2}$. Let us examine them one by one.
In the following, we drop the superscript $2$ from
$F_{n,n}^{2}$ to simplify expressions.

\medskip

\noindent
(1) $F_{2,2}$:

\smallskip

 From \fourtorusintegrand\ and \diagonal, it is easy to read off
the following expression for $F_{2,2}$.
$$ F_{2,2} = 2 \sum_{p_L, p_R}
  \int
\Big( (t+\bar{t})^2 (p_L, \bar{p}_R)^2 - 6
(t+\bar{t}) (p_L, \bar{p}_R) + 6 \Big) \exp(-S)
{d^3 \Omega d^3 \bar{\Omega} \over
[\det {\rm Im} \Omega]^3}  .$$
Since $S$ for $T^2$ is given by \twotorusaction, we can also write
it as
\eqn\twotwo{ F_{2,2} = 2 \Big(
 (t+\bar{t})^2 {\partial^2 \over \partial t^2}
 + 6 (t+\bar{t}) {\partial \over \partial t} + 6 \Big) {\cal Z} ,}
with
\eqn\genustwopartition{ {\cal Z}(t, \bar{t})
= \sum_{p_L, p_R} \int
\exp(-S)  {d^3 \Omega d^3 \bar{\Omega} \over
[\det {\rm Im} \Omega]^3}}
By doing the Poisson resummation in $p_L$ and $p_R$,
one can show that the combination $(t+\bar{t})^2 {\cal Z}(t, \bar{t})$
is invariant under the $T$-duality transformation.
Thus \twotwo\ can also be written as
$$ \eqalign{ F_{2,2} &= {2 \over (t +\bar{t})^2}
{\partial \over \partial t} \Big[
(t +\bar{t})^2  {\partial \over \partial t}
\Big((t+\bar{t})^2 {\cal Z}\Big) \Big] \cr
& = 2 D_t^2 \left[ (t +\bar{t})^2 {\cal Z}( t , \bar{t}) \right] \cr } $$
where $D_t$ is the duality covariant derivative. This means in
particular that the weight of  $F_{2,2} $ is such that
$F_{2,2}(t , \bar{t}) (dt)^2 $
is invariant under the duality transformation.

\medskip

\noindent
(2) $F_{1,1}$:

\smallskip

By extracting a coefficient of $(u_L^1 u_R^1)^3 u_L^2 u_R^2$
from \fourtorusintegrand\ using \diagonal\ for the limit $
T^4 \rightarrow T^2 \times R^2$, we obtain
\eqn\oneone{
\eqalign{ F_{1,1}=
{1 \over 4} \int &
\Big( -(t+\bar{t})^3  (p_L, \bar{p}_R) \Big[
   (p_L,\bar{p}_R)(\bar{p}_L, p_R) -
   (p_L,p_R)(\bar{p}_L, \bar{p}_R) \Big] + \cr
 &~~+  (t+ \bar{t})^2  \Big[ (p_L,\bar{p}_R)^2 +
   5(p_L,\bar{p}_R)(\bar{p}_L, p_R)
-   4 (p_L,p_R)(\bar{p}_L, \bar{p}_R) \Big] - \cr
 &~- (t+\bar{t})\Big[9(p_L, \bar{p}_R)+3(\bar{p}_L, p_R)\Big]
  + 12 \Big) \exp(-S)
{d^3 \Omega d^3 \bar{\Omega}
                   \over [\det {\rm Im} \Omega]^3}\cr}}

This expression is simplified significantly by using
the following formula for the variation of the action \twotorusaction\
with respect to the worldsheet moduli $\Omega_{ab}$.
\eqn\actionderivative{
 {\partial S \over \partial \Omega_{ab}}
   = {i \over 2}(t+\bar{t})
\left( (p_L {1 \over {\rm Im} \Omega})_a
                   (\bar{p}_L {1 \over {\rm Im} \Omega})_b
                 + (p_L {1 \over {\rm Im} \Omega})_b
                   (\bar{p}_L {1 \over {\rm Im} \Omega})_a \right) }
This formula can be derived either by computing the derivative
of $S$ directly or by noting that $(\partial S / \partial {\Omega_{ab}})
 \omega_a \omega_b$ is proportional to an expectation value of
the energy-momentum tensor
$T = g_{i\bar{j}}\partial X^i \partial X^{\bar{j}}$. By using
this formula repeatedly, we are going to reduce $F_{1,1}$ given
by \oneone\ to
\eqn\oneoneone{
\eqalign{ F_{1,1} = &{3 \over 2} \int
 \Big( - (t+\bar{t})(p_L, \bar{p}_R) + 2 \Big) \exp(-S)
 {{d^3 \Omega d^3 \bar{\Omega}} \over [\det {\rm Im} \Omega]^3}
   \cr
 = & {3 \over 2} \left( (t+\bar{t}) {\partial \over \partial t}
               + 2 \right) \int
       \exp(-S)  {d^3 \Omega d^3 \bar{\Omega}
 \over [\det {\rm Im} \Omega]^3} \cr
=& {3 \over 2} \left( (t+\bar{t}) {\partial \over \partial t}
               + 2 \right)
 {\cal Z} = (t +\bar{t})^{-1} \partial_t
\left[ (t + \bar{t})^2{\cal Z} \right] \cr}}
In particular, this shows that $F_{1,1}(t, \bar{t}) (\sqrt{dt})^3
\sqrt{d\bar{t}}$ is invariant under
the duality transformation.

Now let us prove \oneoneone. We first note that the first
term in the integrand of \oneone\ can be rearranged as
$$
\eqalign{
&(p_L, \bar{p}_R) \Big[
   (p_L,\bar{p}_R)(\bar{p}_L, p_R) -
   (p_L,p_R)(\bar{p}_L, \bar{p}_R) \Big] = \cr
&=(p_L, \bar{p}_R) \Big[
   (p_L,\bar{p}_R)(\bar{p}_L, p_R) +
   (p_L,p_R)(\bar{p}_L, \bar{p}_R) \Big] - \cr
&~~~~~~~~~
 -  2(p_L,p_R)(\bar{p}_L, \bar{p}_R)(p_L, \bar{p}_R)  \cr
& =
  {2 \over i (t + \bar{t})}
        \left( {\partial S \over \partial \Omega_{ab}}
             p_{Ra} \bar{p}_{Rb} \right)
     (p_L , \bar{p}_R)
  - {2 \over i(t + \bar{t})}
       \left(   {\partial S \over \partial \Omega_{ab}}
      \bar{p}_{Ra} \bar{p}_{Rb} \right)
     (p_L ,p_R). \cr}
$$
We can then perform the integration-by-parts on ${\cal M}_2$ as
$$
\eqalign{
&\int(p_L, \bar{p}_R) \Big[
   (p_L,\bar{p}_R)(\bar{p}_L, p_R) -
   (p_L,p_R)(\bar{p}_L, \bar{p}_R) \Big]
\exp(-S)  {d^3 \Omega d^3 \bar{\Omega}
 \over [\det {\rm Im} \Omega]^3}
 = \cr
& = {2 \over i (t+\bar{t})}
  \int {\partial \over \partial \Omega_{ab}}
 \left( {
 p_{Ra} \bar{p}_{Rb}
               (p_L ,\bar{p}_R)
-  \bar{p}_{Ra} \bar{p}_{Rb}
               (p_L , p_R) \over [\det {\rm Im} \Omega]^3}
        \right)
  \exp( -S) d^3 \Omega d^3 \bar{\Omega} \cr
& = {1 \over  (t+\bar{t})}
  \int
\Big(  (p_L , \bar{p}_R)^2
 + 4(p_L , \bar{p}_R)
   (\bar{p}_L , p_R)
   -5(p_L , p_R)
   (\bar{p}_L , \bar{p}_R) \Big) \times \cr
&~~~~~~~~~~~\times
\exp(-S)  {d^3 \Omega d^3 \bar{\Omega}
 \over [\det {\rm Im} \Omega]^3}\cr} $$
it is easy to show that there is no contribution from
the boundaries of ${\cal M}_2$.

Substituting this into \oneone , we obtain
\eqn\oneoneoneone{
\eqalign{ F_{1,1}=
{1 \over 4} \int \Big( &   (t+ \bar{t})^2  \Big[
 (p_L,\bar{p}_R)(\bar{p}_L, p_R) +
    (p_L,p_R)(\bar{p}_L, \bar{p}_R) \Big] - \cr
 &- 9(t+\bar{t})(p_L, \bar{p}_R)-3(\bar{p}_L, p_R)
  + 12 \Big) \exp(-S)
{d^3 \Omega d^3 \bar{\Omega}
 \over [\det {\rm Im} \Omega]^3}
\cr}}
We then note
$$
 (p_L,\bar{p}_R)(\bar{p}_L, p_R) +
    (p_L,p_R)(\bar{p}_L, \bar{p}_R)
={2 \over i(t+\bar{t})}
 \left({\partial S \over \partial \Omega_{ab}}
     p_{Ra}\bar{p}_{Rb} \right), $$
and do the integration-by-parts again.
$$
\eqalign{&\int   \Big( (p_L,\bar{p}_R)(\bar{p}_L, p_R) +
    (p_L,p_R)(\bar{p}_L, \bar{p}_R)\Big)\exp(-S)
{d^3 \Omega d^3 \bar{\Omega} \over [\det{\rm Im}\Omega]^3 }
  = \cr
&={2 \over i(t+\bar{t})}
\int {\partial \over \partial \Omega_{ab}}
 \left( {1 \over [ \det {\rm Im} \Omega ]^3}
       p_{Ra}\bar{p}_{Rb} \right)  \exp(-S)
d^3 \Omega d^3 \bar{\Omega} \cr
&={3 \over (t+\bar{t})} \int
  \Big( (p_L, \bar{p}_R) + (\bar{p}_L, p_R) \Big)
 {d^3 \Omega d^3 \bar{\Omega} \over [\det{\rm Im}\Omega]^3}
.\cr }
$$

Substituting this into \oneoneoneone, we recover
\oneoneone\ and this is what we wanted to show.

\medskip

\noindent
(3) $F_{0,0}$:

\smallskip
By extracting a coefficient of $(u_L^1 u_R^1 u_L^2 u_R^2)^2$
from \fourtorusintegrand\ using \diagonal\ for the limit $
T^4 \rightarrow T^2 \times R^2$, we obtain
\eqn\zerozero{
\eqalign{& F_{0,0}=
{1 \over 36} \int
\Big(  (t+\bar{t})^4
  \Big[ (p_L,\bar{p}_R)(\bar{p}_L, p_R) -
   (p_L,p_R)(\bar{p}_L, \bar{p}_R) \Big]^2 + \cr
 & -4 (t+\bar{t})^3  \Big[ (p_L, \bar{p}_R)+
(\bar{p}_L, p_R)\Big] \Big[
   (p_L,\bar{p}_R)(\bar{p}_L, p_R) -
   (p_L,p_R)(\bar{p}_L, \bar{p}_R) \Big] + \cr
 & + (t+ \bar{t})^2  \Big[ 2 (p_L,\bar{p}_R)^2
    + 2 (\bar{p}_L, p_R)^2
  + 40(p_L,\bar{p}_R)(\bar{p}_L, p_R) -
   32 (p_L,p_R)(\bar{p}_L, \bar{p}_R) \Big]  \cr
 &- 36(t+\bar{t}) \Big[ (p_L, \bar{p}_R)+(\bar{p}_L, p_R) \Big]
  + 72 \Big) \exp(-S)
{d^3\Omega d^3\bar{\Omega}
\over [\det {\rm Im}\Omega]^3}\cr}}
As in the case of $F_{1,1}$, we can use the formula
\actionderivative\ to reduce this to
\eqn\zerozerozero{ \eqalign{
F_{0,0}&=\int
\Big( (t+\bar{t})^2(p_L, \bar{p}_R)(\bar{p}_L,p_R)
   -  \cr
&~~~~~~~~-2(t+\bar{t})\Big[
(p_L, \bar{p}_R)+ (\bar{p}_L,p_R)\big]+2 \Big)  \exp(-S)
{d^3 \Omega d^3 \bar{\Omega} \over [\det {\rm Im} \Omega ]^3} \cr
&= \left( (t+\bar{t})^2 {\partial^2 \over \partial t \partial \bar{t}}
      + 2 (t+\bar{t}) \left( {\partial \over \partial t }
+  {\partial \over \partial \bar{t}} \right)
 +2 \right) \int
\exp(-S)  {d^3 \Omega d^3 \bar{\Omega}
 \over [\det {\rm Im}\Omega]^3}\cr
& = \left( (t+\bar{t})^2 {\partial^2 \over \partial t \partial \bar{t}}
      + 2 (t+\bar{t}) \left( {\partial \over \partial t }
+  {\partial \over \partial \bar{t}} \right)
 +2 \right) {\cal Z} =
 \partial_t \partial_{\bar{t}}
\left[ (t+\bar{t})^2 {\cal Z}\right] .\cr}}
Thus in particular $F_{0,0}(t, \bar{t}) dt d\bar{t} $
is invariant under the duality transformation.

Let us prove \zerozerozero. We note that the first term in
the integrand of \zerozero\ can be written as
$$
\eqalign{
&  \Big[ (p_L,\bar{p}_R)(\bar{p}_L, p_R) -
   (p_L,p_R)(\bar{p}_L, \bar{p}_R) \Big]^2 = \cr
&={2 \over i (t + \bar{t})}
\left( {\partial S \over \partial \Omega_{ab}} p_{Ra} \bar{p}_{Rb} \right)
 \Big( (p_L,\bar{p}_R) (\bar{p}_L, p_R)
           + (p_L,p_R)(\bar{p}_L, \bar{p}_R) \Big) - \cr
& ~~~~~~-{4 \over i (t + \bar{t})}
\left( {\partial S \over \partial \Omega_{ab}} p_{Ra} p_{Rb} \right)
  (p_L,\bar{p}_R)(\bar{p}_L, \bar{p}_R)  . \cr}
$$
Therefore
$$
\eqalign{ &
\int \Big( (p_L,\bar{p}_R)(\bar{p}_L, p_R) -
   (p_L,p_R)(\bar{p}_L, \bar{p}_R) \Big)^2 \exp(-S)
 {d^3 \Omega d^3 \bar{\Omega} \over [\det {\rm Im} \Omega]^3}
 = \cr
& = {2 \over i (t + \bar{t})}
\int {\partial \over \partial \Omega_{ab}}
 \Big[
 \Big( p_{Ra} \bar{p}_{Rb}
 \{ (p_L,\bar{p}_R) (\bar{p}_L, p_R)
           + (p_L,p_R)(\bar{p}_L, \bar{p}_R) \} - \cr
&~~~~~~~ -2 p_{Ra} p_{Rb}
  (p_L,\bar{p}_R)(\bar{p}_L, \bar{p}_R) \Big)
 [\det {\rm Im}\Omega]^{-3}\Big]
\exp(-S)d^3 \Omega d^3 \bar{\Omega} \cr
& = {1 \over  (t + \bar{t})}  \int
  \Big( 4 (p_L, \bar{p}_R)^2 (\bar{p}_L, p_R)   +2(\bar{p}_L,p_R)
(p_L,p_R)(\bar{p}_L,\bar{p}_R)- \cr
&~~~~~~~
- 6(p_L, \bar{p}_R)(p_L,p_R)(\bar{p}_L,\bar{p}_R)
         \Big) \exp(-S)
 {d^3 \Omega d^3 \bar{\Omega} \over [\det {\rm Im}\Omega]^3}. \cr}
$$
By substituting this into \zerozero, we obtain
\eqn\zerozerozerozero{
\eqalign{ &F_{0,0}=
{1 \over 36} \int
\Big(  (t+\bar{t})^3
\Big[ 6(p_L,p_R)(\bar{p}_L,\bar{p}_R)(\bar{p}_L,p_R)-\cr
&~~~~~ -
2(p_L,p_R)(\bar{p}_L,\bar{p}_R)(p_L,\bar{p}_R)-
4(p_L,\bar{p}_R)(\bar{p}_L,p_R)^2 \Big] + \cr
&~~~~~ + (t+ \bar{t})^2  \Big[ 2 (p_L,\bar{p}_R)^2
    + 2 (\bar{p}_L, p_R)^2 + \cr
&~~~~~ + 40(p_L,\bar{p}_R)(\bar{p}_L, p_R) -
   32 (p_L,p_R)(\bar{p}_L, \bar{p}_R) \big] - \cr
&~~~~~
 - 36(t+\bar{t}) \{ (p_L, \bar{p}_R)+(\bar{p}_L, p_R) \}
  + 72 \Big) \exp(-S) {d^3 \Omega d^3 \bar{\Omega}
 \over [\det {\rm Im}\Omega]^3}. \cr}}

Next we note that the first term in the integrand of \zerozerozerozero\
can be written as
$$
\eqalign{
&6(p_L,p_R)(\bar{p}_L,\bar{p}_R)(\bar{p}_L,p_R)-
2(p_L,p_R)(\bar{p}_L,\bar{p}_R)(p_L,\bar{p}_R)- \cr
&~~- 4(p_L,\bar{p}_R)(\bar{p}_L,p_R)^2 = \cr
&=
{2 \over i (t+\bar{t})}  {\partial S \over \partial \Omega_{ab}}
     \Big(  -4 p_{Ra} \bar{p}_{Rb}  (\bar{p}_L,p_R) +
 5 p_{Ra} p_{Rb} (\bar{p}_L,\bar{p}_R)- \cr
&~~~~~~~~~~~~~~~~~
{}~~~~~~~~~~~~~~~~~~~ - \bar{p}_{Ra} \bar{p}_{Rb} (p_L,p_R) \Big)  \cr}
$$
Therefore
$$\eqalign{
&\int \Big( 6(p_L,p_R)(\bar{p}_L,\bar{p}_R)(\bar{p}_L,p_R)-
2(p_L,p_R)(\bar{p}_L,\bar{p}_R)(p_L,\bar{p}_R)- \cr
&~~~~~~~~~~~~~~~ - 4(p_L,\bar{p}_R)(\bar{p}_L,p_R)^2 \Big)
\exp(-S) {d^3 \Omega d^3 \bar{\Omega} \over
   [\det {\rm Im} \Omega]^3} =\cr
&= {2 \over i (t+\bar{t})}
  \int {\partial \over \partial \Omega_{ab}}
\Big(
\Big[   -4 p_{Ra} \bar{p}_{Rb}  (\bar{p}_L,p_R) +
5 p_{Ra} p_{Rb} (\bar{p}_L,\bar{p}_R) -\cr
&~~~~~~~~~~~~~~~~~~~~~~~~~~
      - \bar{p}_{Ra} \bar{p}_{Rb} (p_L,p_R)\Big]
 [\det {\rm Im}\Omega]^{-3} \Big) \exp(-S)
d^3 \Omega d^3 \bar{\Omega} \cr
& =
{2 \over  (t+\bar{t})}
\int
\Big( -8(p_L, \bar{p}_R)(\bar{p}_L,p_R)
      +10 (p_L, p_R)(\bar{p}_L,\bar{p}_R) - \cr
&~~~~~~~~~~~~~~~~~~~~~~~~~~ -(\bar{p}_L,p_R)^2
                    - (p_L, \bar{p}_R)^2 \Big) \exp(-S)
 {d^3 \Omega d^3 \bar{\Omega} \over [\det {\rm Im} \Omega]^3}
. \cr}
$$
Substituting this into \zerozerozerozero, we obtain
\eqn\zerozerozerozerozero{
\eqalign{ F_{0,0} =
{1 \over 36} &\int
\Big( (t+ \bar{t})^2  \Big[  24(p_L,\bar{p}_R)(\bar{p}_L, p_R)
   -12 (p_L,p_R)(\bar{p}_L, \bar{p}_R) \Big] - \cr
 &- 36(t+\bar{t}) \Big[ (p_L, \bar{p}_R)+(\bar{p}_L, p_R) \Big]
  + 72 \Big) \exp(-S)
{d^3 \Omega d^3 \bar{\Omega} \over [\det {\rm Im}\Omega]^3}\cr}}

Finally we note that the first term in the integrand of
\zerozerozerozerozero\ can be written as
$$
\eqalign{
&  24(p_L,\bar{p}_R)(\bar{p}_L, p_R)  -
12 (p_L,p_R)(\bar{p}_L, \bar{p}_R)  \cr
&=    36(p_L,\bar{p}_R)(\bar{p}_L,p_R)  -
{24 \over i(t+\bar{t})}
{\partial S \over \partial \Omega_{ab}}
p_{Ra} \bar{p}_{Rb}.\cr}$$
Therefore
$$
\eqalign{
&\int (t+ \bar{t})^2
\Big(  \{  24(p_L,\bar{p}_R)(\bar{p}_L, p_R)
   -12 (p_L,p_R)(\bar{p}_L, \bar{p}_R) \Big)
\exp(-S) {d^3 \Omega d^3 \bar{\Omega}
 \over [\det {\rm Im}\Omega]^3} = \cr
& =
36\int
\Big((t+\bar{t})^2 (p_L, \bar{p}_R)(\bar{p}_L,p_R)
   - (t+\bar{t}) \{(p_L, \bar{p}_R)+ (\bar{p}_L,p_R)\} \Big)
\times \cr &~~~~~~~~~~~~~~~~\times \exp(-S)
 {d^3 \Omega d^3 \bar{\Omega} \over [\det {\rm Im}\Omega]^3}.\cr}
$$

Substituting this into \zerozerozerozerozero, we find
that $F_{0,0}$ is expressed as \zerozerozero, and this
is what we wanted to show.

\subsec{Genus Two; Evaluation II}

We have shown that, for $T^2 \times R^2$,
the $N=2$ string amplitudes at genus $2$ are given by
\eqn\genustwoamplitude{ \eqalign{
F_{2,2} & = 2 (t+ \bar{t})^2 D_t^2 {\cal Z} \cr
F_{1,1} & = {3 \over 2}  (t+ \bar{t}) D_t {\cal Z} \cr
F_{0,0} & = (t+ \bar{t})^2 D_t D_{\bar{t}} {\cal Z} \cr} }
where
$$ {\cal Z} = \sum_{p_L, p_R} \int_{{\cal M}_2} \exp(-S(p_L, p_R))
{d^3 \Omega d^3 \bar{\Omega} \over [\det {\rm Im} \Omega]^3 }. $$
We shall see that these expressions are consistent
with the harmonicity equation.

To understand $F_{n,n}$ better, we shall first prove the
following two key properties of ${\cal Z}$;

\noindent
(1) ${\cal Z}$ is a sum of two terms, one
depends only on the K\"ahler moduli $\sigma = (8\pi i)^{-1} t$ and
$\bar{\sigma}$ and another
depends only on the complex moduli $\rho$ and $\bar{\rho}$
up to a factor $({\rm Im \sigma})^{-2}$.
\eqn\keyone{ {\cal Z} = f(\sigma, \bar{\sigma}) + ({\rm Im} \sigma)^{-2}
          \tilde{f}(\rho, \bar{\rho}) }

\noindent
(2) $f$ and $\tilde{f}$ are eigen-functions of Laplacians
on the K\"ahler and the complex moduli spaces respectively.
\eqn\keytwo{ \eqalign{ \partial_t \partial_{\bar{t}}
         \left[ (t + \bar{t})^2 f \right]& =
         2 f \cr
        4({\rm Im} \rho)^2 \partial_\rho \partial_{\bar{\rho}}
               \tilde{f} &=
              2 \tilde{f} \cr} }

These properties, combined with the large $t$ behavior of ${\cal Z}$,
$$ {\cal Z} \rightarrow \int
{d^3 \Omega d^3 \bar{\Omega} \over [\det {\rm Im} \Omega]^3}
\sim \int_{{\cal M}_2} (c_1)^3~~~~(t, \bar{t} \rightarrow \infty) $$
where $c_1$ is the first Chern class of the Hodge bundle
over the moduli space ${\cal M}_2$
and the mirror symmetry $\sigma \leftrightarrow \rho$,
completely determines $f(\sigma,\bar{\sigma})$
and $\tilde{f}(\rho,\bar{\rho})$
as
$$ \eqalign{ f(\sigma,\bar{\sigma}) & =
  \sum_{n,m}{1 \over (n+m\sigma)^2(n+m\bar{\sigma})^2} \cr
 \tilde{f}(\rho,\bar{\rho}) & =
  \sum_{n,m}{({\rm Im}\rho)^2 \over
            (n+m\rho)(n+m\bar{\rho})}. \cr} $$
By substituting this into \genustwoamplitude\ and \keyone,
we obtain the following expression for $F_{n,n}$.
\eqn\genustwoexpression{ \eqalign{
F_{0,0} & = 2f
=\sum_{n,m}{2 \over (n+m\sigma)^2(n+m\bar{\sigma})^2} \cr
F_{1,1} & = {3\over 2}(t+\bar{t})D_t f
 = \sum_{n,m}{3 \over (n+m\sigma)^3(n+m\bar{\sigma})} \cr
F_{2,2} & = 2(t+\bar{t})^2 D_t^2 f =
\sum_{n,m}{12 \over (n+m\sigma)^4} \cr} }
In particular, $F_{2,2}$ is holomorphic in $\sigma$
$$  \partial_{\bar{\sigma}} F_{2,2} = 0 $$
and is given by the Eisenstein series of degree 4.
These expressions for $F_{n,n}$ are combined nicely
as
$$ \eqalign{ F(u_L,u_R) & =
 \sum_{n=-2}^2 {4 \choose 2 + n}^2 F_{n,n}\ (u^1_Lu^1_R)^{2+n}
(u^2_Lu^2_R)^{2-n}
\cr
& = 12 \sum_{(n,m)\neq (0,0)}
     \left( {u^1_L u^1_R \over n + m \sigma} +
       {u^2_L u^2_R \over n + m \bar{\sigma}} \right)^4 .\cr} $$

Now let us prove \keyone\ and \keytwo. We will use
$$ \eqalign{
& \partial_\rho (p_L , \bar{p}_R)  =
 \partial_\rho (\bar{p}_L , p_R) =
 {i \over 2 {\rm Im}\rho}(\bar{p}_L, \bar{p}_R) \cr
& \partial_\rho \Big[ {1 \over {\rm Im} \rho} (p_L , p_R)\Big]
={i \over 2({\rm Im} \rho)^2}
 \Big( (p_L, \bar{p}_R) + (\bar{p}_L, p_R) \Big) \cr
&\partial_\rho \Big[ {\rm Im} \rho (\bar{p}_L , \bar{p}_R)\Big]
=0\cr} $$
and
$$ \partial_\rho S = {i(t+\bar{t}) \over 2 {\rm Im} \rho}
(\bar{p}_L, \bar{p}_R) ,$$
which follows from the definition of $(p_L, \bar{p}_R)$
etc. Therefore
$$ \eqalign{
\partial_\rho D_t & {\cal Z}  =
\partial_\rho \int
\Big( -(p_L, \bar{p}_R) + {2 \over (t +\bar{t})} \Big)
\exp(-S) {d^3 \Omega d^3 \bar{\Omega} \over [\det {\rm Im}
\Omega ]^3 } \cr
& =
\int \Big( {i(t+\bar{t}) \over 2 {\rm Im} \rho}
 (\bar{p}_L, \bar{p}_R) (p_L, \bar{p}_R)
- {3i \over 2 {\rm Im} \rho}(\bar{p}_L, \bar{p}_R)  \Big)
\exp(-S) {d^3 \Omega d^3 \bar{\Omega} \over [\det {\rm Im}
\Omega ]^3 } .\cr} $$
We then note
$$ i (t+\bar{t}) (\bar{p}_L, \bar{p}_R) (p_L, \bar{p}_R)
   = {1 \over {\rm Im} \rho}
 \bar{p}_{La} \bar{p}_{Rb}
      {\partial S \over \partial \Omega_{ab}} .$$
We can then perform integration-by-parts to obtain
$$ \eqalign{
&\int {i(t+\bar{t}) \over 2 {\rm Im} \rho}
 (\bar{p}_L, \bar{p}_R) (p_L, \bar{p}_R)
\exp(-S) {d^3 \Omega d^3 \bar{\Omega} \over [\det {\rm Im}
\Omega ]^3} = \cr
& = \int {-1 \over 2 ({\rm Im} \rho)^2}
   {\partial \over \partial \Omega_{ab}}\left( {
\bar{p}_{La} \bar{p}_{Rb} \over [ \det {\rm Im} \Omega ]^3}\right)
\exp(-S) d^3 \Omega d^3 \bar{\Omega} \cr
& = -{3i \over 2 {\rm Im} \rho}\int
 (\bar{p}_L, \bar{p}_R)
\exp(-S) {d^3 \Omega d^3 \bar{\Omega} \over [\det {\rm Im}
\Omega ]^3 } .\cr}$$
Thus we found
$$ \partial_\rho D_\sigma {\cal Z} =
   ({\rm Im} \sigma)^{-2} \partial_\rho \partial_\sigma
  \Big[ ({\rm Im}\sigma)^2 {\cal Z} \Big] = 0 .$$
Similarly we can prove
$$\partial_\rho \partial_{\bar{\sigma}}\Big[
({\rm Im}\sigma)^2 {\cal Z}
\Big]=
\partial_{\bar{\rho}} \partial_{\sigma}\Big[
({\rm Im}\sigma)^2 {\cal Z}
\Big]
 = \partial_{\bar{\rho}} \partial_{\bar{\sigma}}\Big[
({\rm Im}\sigma)^2 {\cal Z}
\Big]
 = 0 .$$
Therefore ${\cal Z}$ is a sum of $f(t, \bar{t})$
and $(t+\bar{t})^{-2} \tilde{f}(\rho, \bar{\rho})$ as
in \keyone.

To prove \keytwo, we first compute
$$ \eqalign{
& \Big( (t +\bar{t})^2 D_t D_{\bar{t}} +
      4 ({\rm Im} \rho)^2 \partial_\rho \partial_{\bar{\rho}}
\Big) {\cal Z} = \cr
&= \int \Big(
 (t +\bar{t})^2 \left[ (p_L, \bar{p}_R)(\bar{p}_L, p_R)
        + (p_L,p_R)(\bar{p}_L,\bar{p}_R) \right] - \cr
&~~~~~~- 3 (t+\bar{t}) \left[ (p_L,\bar{p}_R) +
   (\bar{p}_L,p_R) \right] + 2 \Big) \exp(-S) {d^3 \Omega
d^3 \bar{\Omega} \over [\det {\rm Im}\Omega]^3 } .\cr}$$
By using
$$ (t +\bar{t})^2 \left[ (p_L, \bar{p}_R)(\bar{p}_L, p_R)
        + (p_L,p_R)(\bar{p}_L,\bar{p}_R) \right]
= -2i(t +\bar{t})p_{Ra} \bar{p}_{Rb} {\partial S \over
 \partial \Omega_{ab}}, $$
one can show
$$ \eqalign{ \int \Big(&
 (t +\bar{t})^2 \left[ (p_L, \bar{p}_R)(\bar{p}_L, p_R)
        + (p_L,p_R)(\bar{p}_L,\bar{p}_R) \right] - \cr
& - 3 (t+\bar{t}) \left[ (p_L,\bar{p}_R)
+(\bar{p}_L, p_R) \right] \Big) \exp(-S) {d^3 \Omega
d^3 \bar{\Omega} \over [\det {\rm Im}\Omega]^3 }
= 0 \cr}$$
by integration-by-parts. Thus ${\cal Z}$ is an eigen-function
of a Laplacian
$$(t +\bar{t})^2 D_t D_{\bar{t}} +
4 ({\rm Im} \rho)^2 \partial_\rho \partial_{\bar{\rho}}
= 4\Big( ({\rm Im} \sigma)^2 D_\sigma D_{\bar \sigma}
+  ({\rm Im} \rho)^2 \partial_\rho \partial_{\bar{\rho}} \Big)$$
as
$$ \Big( ({\rm Im} \sigma)^2 D_\sigma D_{\bar \sigma} +
      ({\rm Im} \rho)^2 \partial_\rho \partial_{\bar{\rho}}
\Big) {\cal Z} = 2 {\cal Z} .$$
Since ${\cal Z}$ is a sum of $f(\sigma,\bar{\sigma})$ and $
({\rm Im}\sigma)^{-2}\tilde{f}(\rho,\bar{\rho})$ as in \keyone,
this means that $f$ and $\tilde{f}$ are also eigen-functions
of $({\rm Im}\sigma)^2 D_\sigma D_{\bar{\sigma}}$ and
$({\rm Im} \rho)^2 \partial_\rho
\partial_{\bar{\rho}}$ as in \keytwo, and this is what we wanted
to show.

\subsec{Topological Interpretation at $g=2$}

We have found that $F_{2,2}$ is holomorphic in $t$ and
is given by the Eisenstein series of degree $4$.
The holomorphicity of $F_{2,2}$ implies
\ref\wittentop{E. Witten, ``Topological Mirrors and Quantum Ring,''
in {\it Essays on Mirror Manifolds,} ed. by S.-T. Yau
(International Press, Hong-Kong 1992).},\holan\
that $F_{2,2}$ should ``count'' the number of holomorphic
maps from genus-$2$ Riemann surfaces to $T^2$.

Since $F_{2,2}$ is independent of $\bar{t}$, let us
regard $t$ and $\bar{t}$ to be independent and
take $\bar{t} \rightarrow \infty$ limit in \twotwo\ while
keeping $t$ to be finite. This limit imposes constraint
on the period matrix $\Omega_{ab}$ as
\eqn\constraint{ \Omega_{ab} (m^b + \rho r^b) = (n_a + \rho s_a)  . }
In this case, the map $X : \Sigma \rightarrow T^2$
characterized by the string momenta $p_L, p_R$ become
a holomorphic map.
There are $2$ equations for $3$ independent components of
$\Omega_{ab}$ constraints. Thus a solution to the constraint
should be parametrized by one complex parameter.
It is easy to write down the most general solution.
Since $\Omega_{ab}$ is symmetric, we can parametrize it
by 3 complex parameters $u,v,w$ as
$$ \eqalign{
\Omega_{ab}  = &  u (Im \Omega \bar{\alpha})_a
(Im \Omega \bar{\alpha})_b + \cr
  &   + v \left[ \epsilon_{ac} \alpha^c (Im \Omega \bar{\alpha})_b
         + (Im \Omega \bar{\alpha})_a \epsilon_{bc} \alpha^c \right]
    + w \epsilon_{ac} \alpha^c \epsilon_{bd} \alpha^d, \cr}$$
where $\alpha^a = m^a + \rho r^a$. For fixed $u,v,w$,
this is a non-linear equation since $Im \Omega$ in the right
hand side also depends on $u,v,w$.  This however will
not cause complication later since the values of $u$ and $v$ are
fixed by the constraints and the dependence on $w$
turns out to be simple as we shall see. In this parametrization,
the solutions to the constraints correspond to
$$
\eqalign{& u = u_0
= (\alpha Im \Omega \bar{\alpha})^{-2}  (n_a + \rho s_a) \alpha^a \cr
         & v = v_0 = (\alpha Im \Omega \bar{\alpha})^{-2}  (n_a + \rho s_a)
                \epsilon^{ab} (Im \Omega \bar{p})_b \cr} $$
and $w$ is arbitrary.
The term of the action which blows up
in the $\bar{t} \rightarrow \infty$ limit is now
of the form
$$
\eqalign{ \bar{t} (\bar{p}_L, p_R) = {\bar{t} \over {\rm Im} \rho}
(\alpha Im \Omega \bar{\alpha})^3
        \big( |u-u_0|^2 + (\det Im \Omega)^{-1}
                |v-v_0|^2 \big).\cr}
$$
The exponentiated action becomes in this limit
$$ \eqalign{ &\exp(-S) \sim \cr
& \sim  \left( { {\rm Im} \rho \over \bar{t}} \right)^2
[\alpha Im \Omega \bar{\alpha}]^{-6} [\det Im \Omega]
 \delta^{(2)}(u-u_0) \delta^{(2)}(v-v_0)  \exp(-t(p_L, \bar{p}_R)).
\cr} $$
Since $\Omega_{ab} \alpha^b = (n_a + \rho s_a)$, it follows
$$ \eqalign{ (p_L, \bar{p}_R) & = 4{
     \alpha^a {\rm Im} \Omega_{ab} \bar{\alpha}^b
 \over {\rm Im} \rho}\cr
& = 4 {
{\rm Im} ( \alpha^a  \Omega_{ab} \bar{\alpha}^b )
\over {\rm Im} \rho}\cr
& = 4 (s_am^a - r_a n^a), \cr} $$
namely $(p_L, \bar{p}_R)$ is a degree of the holomorphic map
from $\Sigma$ to $T^2$.
It is convenient to change the integration variables from
$\Omega_{ab}$ to $u,v,w$. The Jacobian is easily computed as
$$ d^3 \Omega d^3 \bar{\Omega} =
     (\alpha Im \Omega \bar{\alpha})^6 d^2u d^2v d^2w $$
Thus $(p Im \Omega \bar{p})^{-6}$ from the exponentiated action
cancels with the Jacobian.

To compute $F_{2,2}$, we need to apply $D_t^2$ on ${\cal Z}$ as
in \twotwo. In the $\bar{t} \rightarrow \infty$ limit,
$D_t^2$ acting on $\exp(-S)$ reduces to $\partial_t^2$, and
the integrand for $F_{2,2}$ becomes
$$
\left( { p Im \Omega(w) \bar{p} \over \det {\rm Im} \Omega(w)}
\right)^2 \exp[2 \pi i \sigma (s_am^a - r_a n^a)]
d^2 w $$
where $\sigma = (8 \pi i)^{-1} t$ and
$$ \Omega_{ab}(w) = \Omega^{0}_{ab}
         + w \epsilon_{ac} \alpha^c \epsilon_{bd} \alpha^d , $$
with $\Omega^{0}_{ab}$ being a special solution to
the constraint \constraint .

Since
$$ {\partial \over \partial w} \Omega_{ab}(w) =
 \epsilon_{ac} \alpha^c \epsilon_{bd} \alpha^d, $$
we can write
$$\eqalign{
 \left( { p Im \Omega(w) \bar{p} \over \det {\rm Im} \Omega(w)}
\right)^2 & =
 ({\rm Im} \Omega^{-1})^{ab} \partial_w \Omega_{bc}(w)
 ({\rm Im} \Omega^{-1})^{cd} \partial_{\bar{w}} \bar{\Omega}_{da}
  (\bar{w}) \cr
& = \partial_w \partial_{\bar{w}} {\rm trace} \log {\rm Im}
 \Omega \cr
& = \partial_w \partial_{\bar{w}} \log \det {\rm Im} \Omega \cr}$$
Thus we can interpret that $F_{2,2}$ computes the first Chern class
of the Hodge bundle over the one dimensional moduli space of
holomorphic maps from $\Sigma$ to $T^2$. In section 5, we will
further elaborate on this point and show that we can
reproduce the Eisenstein series of degree $4$ from this
topological point of view.

\subsec{Harmonicity Equation on $T^2 \times R^2$}

Now that we have the explicit expression \genustwoexpression\
for $F_{n,n}$, we would like to check whether the harmonicity
equations \harmonicL\ and \harmonicR\ are consistent with it.

Let us first write down the harmonicity equation \harmonicL\
on $T^2 \times R^2$ for general value of $g$.
In terms of the components, the equation is
\eqn\anomalycomponent{
\eqalign{ D_{t^{22}} F^{g}_{n,m} &- D_{t^{12}} F^{g}_{n-1,m}
+ \cr
&+{2g-2+m \over 2g-2-m+1} \Big(
 D_{t^{21}} F_{n,m-1}^{g} - D_{t^{11}} F_{n-1,m-1}^{g}
\Big) = 0 \cr} .}

Suppose $t^{22}$ couples to the marginal operator
$\partial_z X^{\bar{1}} \partial_{\bar{z}} X^{1}$
where $X^1$ is the coordinate in the $T^2$ direction,
namely $t^{22} = \bar{t}$ in the notation in this section.
In this case, $t^{12}$, $t^{21}$ and $t^{11}$ couple to
$\partial_z X^{2} \partial_{\bar{z}} X^{1}$,
$\partial_z X^{\bar{1}} \partial_{\bar{z}} X^{\bar{2}}$ and
$\partial_z X^{2} \partial_{\bar{z}} X^{\bar{2}}$ respectively.
In this case, it is easy to see that the only nontrivial
case in \anomalycomponent\ is when $n=m$, otherwise
each term in the equation vanishes identically.
Since $X^2$ is in the $R^2$ direction, $X^2(z,\bar{z})$ is a single
valued function on the Riemann surface $\Sigma$. It is then
straightforward to compute insertions of these operators
in $F^{g}$ and obtain
$$ \eqalign{
& (t +\bar{t}) D_{t^{12}} F^{g}_{n-1,n} = (2g-2+n) F^{g}_{n-1,n-1} \cr
& (t +\bar{t})D_{t^{21}} F^{g}_{n,n-1} = (2g-2+n) F^{g}_{n-1,n-1} \cr
& (t +\bar{t})D_{t^{11}} F^{g}_{n,n} = (g+n) F^{g}_{n,n} \cr} $$
We can derive these formula by writing, for example,
$\partial_z X^{2} \partial_{1} X^{1} =
\partial_z \Big( X^{2} \partial_{1} X^{1} \Big)$ and by doing
integration-by-parts.
By substituting them into \anomalycomponent, we obtain
\eqn\anomalytbar{
   (t +\bar{t}) D_{\bar{t}} F_{n,n}^{g} = {2g-2+n\over 2g-2-n+1}
   (g-n) F_{n-1,n-1}^{g} }
when $t^{22} = \bar{t}$. Similarly when $t^{11}=t$,
\anomalycomponent\ becomes
\eqn\anomalyt{  (t +\bar{t})
D_{t} F_{n,n}^{g} = {2g-2-n \over 2g-2+n+1}
   (g+n) F_{n+1,n+1}^{g} }
By combining these two equations, we also find that
$F_{n,n}^{g}$ is an eigen-function of the Laplace operator
\eqn\laplaceq{(t +\bar{t})^2 D_t D_{\bar{t}} F^{g}_{n,n} =
(g-n)(g+n-1) F_{n,n}^{g} .}

When $g=2$, the holomorphic anomaly equations, \anomalytbar\
and \anomalyt, gives
$$ \eqalign{
& D_{\bar{t}} F_{1,1} = {3 \over 2} F_{0,0} \cr
& D_t F_{1,1} = {3 \over 4} F_{2,2} \cr
& D_t F_{0,0} = {4 \over 3} F_{1,1} .\cr}
$$
It is straightforward to check that,
combined with the Laplace equation \laplaceq,
they are consistent with the explicit expressions
\genustwoexpression\ for $F_{n,n}^2$.
Now we can apply the harmonicity equations, \anomalytbar\
and \anomalyt, to compute $F^g$ for all $g$.

\subsec{$g \geq 3$}

We have verified that the harmonicity equations
\anomalytbar\ and \anomalyt\ are consistent with the
explicit computation at genus $2$. Let us now use
the harmonicity equations to determine $F^{g}$ for
all $g \geq 3$. The two equations imply
$$ \eqalign{
(t +\bar{t})^2 D_t D_{\bar{t}} F^{g}_{n,n} &=
(g-n)(g+n-1) F_{n,n}^{g} \cr
(t +\bar{t})^2 D_{\bar{t}}D_t F^{g}_{n,n} &=
(g+n)(g-n+1) F_{n,n}^{g} ,\cr}$$
and therefore
$$ [D_t , D_{\bar{t}}] F^{g}_{n,n}=
-2g(t +\bar{t})^{-2} F^{g}_{n,n} .$$
Combined with the hermiticity condition
$\overline{F^{g}_{n,n}} = F^{g}_{-n,-n}$, we find
$F_{n,n}^{g}$ $(\sqrt{dt})^{(g+n)}$ $(\sqrt{d\bar{t}})^{(g-n)}$
is invariant under the duality transformation.

Using the fact that
$${\tilde F}^g_{0,0}=F^g_{0,0}(t+\bar t)^g$$
is a modular function of weight
zero, and that it is an eigenstate of Laplacian \laplaceq\
and that as $t\rightarrow \infty$ it can at most have power law
singularity in $t+\bar t$ (as it is becoming equivalent to $R^4$)
allows us to solve for it (up to an overall constant).
In particular we learn from \laplaceq\ that for large $t$
$${\tilde F}^g_{0,0}\sim (t+\bar t)^g$$
Now using the modular invariance we can get the rest
by acting with $SL(2,Z)$ (note that $SL(2,Z)$ transformations
commute with the Laplace operator and so will give you
another function with the same eigenvalue for Laplace operator)\foot{
We are thankful to A. Lesniewski for discussion on this point.}.
We thus learn, in this way, that
\laplaceq\ has a solution for $F_{0,0}^{g}$ as
$$ F_{0,0}^{g} = ({\rm const}) \times
\sum_{(n,m) \neq (0,0)} {1 \over |n+m\sigma|^{2g}}. $$
That this solution is unique follows from the fact that
if we had another solution, by subtracting the two solutions we
get a function which vanishes at infinity--this means that it
is the eigenstate of Laplacian with a positive eigenvalue,
which is the wrong sign.  Thus we have a unique solution.
Note that the constant appearing in front of our solution
cannot depend on the complex structure of $T^2$ because we can always
take $t\rightarrow \infty$ in which case the answer will be the
partition function on $R^4$ which clearly is independent of
which complex structure we chose for $T^2$ before blowing it up.

We can now use \anomalytbar\ and \anomalyt\ to compute
the rest of $F_{n,n}^{g}$ to obtain
$$ F^g(u_L,u_R) = ({\rm const}) \times
\sum_{(n,m) \neq (0,0)}
|n+m\sigma|^{2g-4}\left({u_L^1u_R^1 \over n+m\sigma}
+{u_L^2u_R^2 \over n+m\bar{\sigma}} \right)^{4g-4} .$$

\newsec{Topological Interpretation of $N=2$ String Amplitudes on $T^2\times
R^2$}

Having seen that the genus 1 and genus 2 computation of $N=2$ string
amplitudes on $T^2\times R^2$
have a topological interpretation we now ask the same
question in all genera.  In the general case we use the result
discussed in section 2, and in particular apply equation \topeq\
to our special case where target space is $T^2\times R^2$.

Let us recall equation \topeq :
$$F^g_{2g-2,2g-2}\big|_{\bar t \rightarrow \infty}=
 \int_{{\cal M}^g}
k^{g-1}\wedge c_{n-1}({\cal V})\wedge
{\cal J}$$
This equation shows that the top instanton number amplitude, in the limit
$\bar t \rightarrow \infty$ can be reinterpreted topologically
by doing a topological computation on the moduli space of holomorphic
maps.  Let us see how this works.
In the case of genus one the above computation
is exactly the same as counting the holomorphic maps from torus to torus,
because the ${\cal J}$ insertion precisely absorbs the zero mode in the
direction of $R^2$ and so we are back to counting holomorphic maps
from genus one to genus one, which was done in \holan .

For genus $g$, the moduli of
holomorphic maps ${\cal M}$ has dimension $(2g-2+1)$ for degree
bigger than zero.  This corresponds to double
covering of the torus by the Riemann surface having
$(2g-2)$ branch points and $(+1)$ comes from choice of the $R^2$
coordinate of the holomorphic map.  Note that all holomorphic
maps to $T^2\times R^2$ will lead to constant maps as far as the
$R^2$ factor is concerned.  Thus
pulling back the volume form $V$ and integrating over the Riemann
surface will lead us to the statement that $k$ is precisely
the $(1,1)$ form on $\cal M$ in the direction of changing the $R^2$ image.
The
bundle ${\cal V}$ in our case is the same as holomorphic one forms,
simply because the normal bundle is simply the $R^2$ direction (i.e.
the fermion zero modes in the $R^2$ direction).  In other words
 ${\cal V}$ is simply the Hodge bundle ${\cal H}$ on the moduli
of genus $g$ surfaces, restricted in our case to the moduli
of Riemann surfaces which holomorphically cover a fixed torus.
The top chern class is $g$, but we are instructed to take
the $(top-1)$ class, which is $c_{g-1}({\cal H})$. Note that the
dimension of ${\cal M}$ agrees with $(g-1)+(g-1)+1$ as expected
from \topeq .  Let us first consider the case of $g=2$.
In this case we are instructed to compute $\int k\wedge c_1\wedge {\cal J}$
over
the moduli space of holomorphic maps which is of dimension $3$; 2 coming
from the choices of two branch points and 1 from the image of the
map on $R^2$.  As discussed above the $k$ integrates over the $R^2$ part
and gives the volume in the $R^2$ direction.  Moreover ${\cal J}$
gives the volume form over the torus, i.e. absorbs the zero mode
corresponding to shift of the origins of the map on the torus direction.
Note that if we did not have ${\cal J}$ and if we have $c_2$
instead of $c_1$ the computation would have been the standard
$N=2$ topological computation which would have vanished
because of the flatness of the torus.  This agrees with the
general argument that the $J$ insertion is crucial
for a non-vanishing answer.
We are thus left with $\int c_1$ over the moduli of holomorphic
maps from genus 2 to torus, up to a shift in the origin of the torus.
This is precisely the object we encountered in explicit computation
in section 3.

\subsec{Genus 2 Topological Computation}
We have seen that the $\bar t \rightarrow \infty$ of
genus 2 computation of the top
component amplitude  is the same as integration of
the first Chern class $c_1$ of the Hodge bundle
over the one dimensional space of moduli of holomorphic
maps.  Moreover using other argument we have shown that the top
component is proportional to $E_4$.  We will now prove that the
answer being
proportional to $E_4$ could have also been derived using the
direct topological computation.

To this end we have to use the fact that $c_1$ for genus
2 can be written as
$$ c_1= {2\pi i}\partial \overline \partial {\rm log} {\rm det}
{\rm Im} \Omega$$
and try to use integration by parts to integrate over moduli
of holomorphic curves.  However in order to do this
we cannot directly use the above expression because ${\rm det}\
{\rm Im}\Omega$ is not a modular invariant object.  Instead
we write it as\foot{Which is the same trick that give
the 2 loop bosonic string amplitude \ref\mokn{G. Moore,``
Modular Forms and Two Loop String Physics,'' Phys. Lett. 176B (1986) 369
\semi A. Belavin and V. Knizhnik, A. Morozov and A. Perelomov,
``Two and Three Loop Amplitudes in the Bosonic String Theory,''
Phys. Lett. 177B (1986) 324.}}.
$$c_1={1\over 2\pi i}\partial \overline \partial {\rm log}
\Big[ {\rm det}
{\rm Im} \Omega \big(\prod_{even\ \theta \ functions}\vartheta
\bar \vartheta \big)^{1/5} \Big]$$
which is modular invariant.
Note that product of even $\theta$ functions has no zeroes
in the interior of the moduli space for $g=2$ (a fact that fails to be true
for higher genera).  Since we have a total derivative we
can integrate by parts and we thus come to the point on the
moduli of holomorphic maps which corresponds either to a
handle degeneration or to splitting to two genus 1 curves.
The product of even theta functions in the handle degeneration
case has a zero of the order $z^{1/2}$ and in the case of
splitting a zero of the order $z$.
So in order to compute
$\int_{\cal M}c_1$ we simply have to count
how many holomorphic curves exist which go from a handle degenerated
genus 2 to torus and multiply it by $1/10$ and add to it the number
of holomorphic curves which exist when we have the splitting
case and multiply it by $1/5$.  This is described mathematically
by the statement that
\eqn\useit{c_1={1\over 10}(2\delta_1 +\delta_0)}
where $\delta_1$ denotes the first chern class of a bundle
whose divisor is the boundary of moduli space corresponding
to genus 2 splitting to two genus 1 curves and $\delta_0$ denotes
the the corresponding one where the divisor is the boundary of moduli
space where the genus 2 curve has a handle degeneration.  Note that
we have chosen coordinates on the moduli space such that a symmetry
factor of $1/2$ in the $\delta_0$ and $\delta_1$ degenerations are included.

Using \useit\ we are in a position to compute the
genus 2 topological amplitude in terms of genus 1 amplitude\foot{We are
grateful to R. Dijkgraaf for explaining this to us.}.
First note that a genus 2 covering of a torus will lead to two
branch points.  The degenerate genus 2 curves can occur only
when the two branch points collide.  Not every colliding
branch points give rise to degenerate Riemann surfaces,
as some of them simply convert 2 branch points of order 2
to a single one of order three.  Those would not contribute
to our amplitude.  To count the degenerations of the other type,
note that if you remove the degenerate preimage we end up in the
handle degeneration case to a holomorphic map from torus
to torus where we have marked two of the covering sheets (the
ones which get glued over the handle degeneration)
and in the splitting case to two genus one curves
connected by a tube.  In the handle degeneration case if the
remaining genus 1 to torus map is of degree $n$, we have $n(n-1)/2$
ways to choose the sheets, and so putting all the contributions
of these together, and denoting the genus 1 answer by $F_1$
(the topological part of it which is $dF_1/dt=\eta'/2\pi i\eta$)
we see that the handle degeneration gives (noting that
$1/2$ is already counted in the definition of $\delta_0$)
$${1\over 10}\cdot \big[{d^2F_1\over dt^2}-({dF_1\over dt}
+{1\over 24})\big]$$
(note that each $d/dt$ gives a factor of $n$--note that since
we are in the topological limit of $\bar t \rightarrow \infty$
we do not have covariantization of $d/dt$).  We have added
$+1/24$ to $dF_1/dt$ to eliminate to degree zero part of the map
which we take into account separately below.
 Similarly when
we get the splitting case we get two maps from two different
genus 1 curves to our torus.  We simply have to choose a sheet
from each one to identify with the other.  If one of them
is a covering of order $n$ and the other of order $m$, we get
$nm$ ways of doing this.  We also have to divide by a symmetry
factor of $1/2$ because of the $Z_2$ symmetry of exchanging
the two genus 1 curves.  We thus get a contribution from
the splitting case (noting that the symmetry
factor $1/2$ is already included
in the definition of $\delta_1$)
$${1\over 5}\cdot \big( {dF_1\over dt}+{1\over 24}\big)^2$$
In addition to these two contributions we have bubbling type
contributions, which correspond to degenerate maps from a genus
2 to the torus, where the genus 2 curve is itself a torus glued
to another torus, where one torus gets mapped to a constant, and
the other gets holomorphically mapped to the torus.  The $c_1$
of this family will simply be the $c_1$ of the genus 1 curves times
the one point function of the genus 1 answer.  Since $c_1$ on
the genus 1 moduli space gives 1/12, the bubbling contribution
is given by
$${1\over 12}({dF_1\over dt}+{1\over 24})$$
There is also going to be an overall constant contribution
coming from genus 2 curves which map to a constant.
Using the topological formula \topeq , and the fact that in this
case $\int k\wedge {\cal J}$ absorb the volume integral over
$T^2\times R^2$, this should be
$c_3({\cal H}\oplus
{{\cal H}})=2c_1({\cal H})c_2({\cal H})$ and using the fact that
$2c_2=(c_1)^2$ it is given by $(c_1)^3$.  Integrated
over moduli of genus 2 curves, this gives ${1\over 2880}$.
Putting all these three contributions together we find
$${1\over 10}\cdot\big[{d^2F_1\over dt^2}-({dF_1\over dt}
+{1\over 24})\big]+
{1\over 5}\cdot \big( {dF_1\over dt}+{1\over 24}\big)^2+
{1\over 12}({dF_1\over dt}+{1\over 24})+{1\over 2880}$$

It is quite miraculous that all the terms which are not
second order in derivatives of $t$ disappear as they should
in order to end up with a function of a definite modular weight.
Moreover $E_4$ which was shown to be proportional to the
genus 2 answer is proportional to ${d^2F_1\over dt^2}+2\big[{dF_1\over dt}
\big]^2$
as expected.  We thus learn that
$$F^2_{2,2}={1 \over 2880}
E_4={1\over 10}({d^2F_1\over dt^2}+2({dF_1\over dt})^2)$$

\subsec{$g \geq 3$}

If we consider  $g \geq 3$ the above topological computation formally vanishes,
because we get a higher power of $k$.  Since all of them are
in the direction of $R^2$, and there is only one such direction
on the moduli space and if the topological amplitude
were give by the above formula, we would get zero.  In fact
this is precisely an example of the type mentioned at the end
of section 2, where the extra insertions go to modifying the bundle
${\cal V}$.  This is clear from the explicit attempt in computation
of the amplitude for $g \geq 3$ because then we can no longer replace
the fermion fields by the zero mode wave functions,
as is possible for genus 2--some of the fermions
are contracted, giving us Greens functions, which are reinterpreted
as curvature of a bundle, as in \wiwzw .
In such a case presumably methods similar to those of \wiwzw\ should
be applicable to determine the new bundle $\widetilde{\cal V}$ which
we expect to be of rank $2g-2$, and for which the amplitude can be written as
$$F^g_{2g-2,2g-2}\big|_{\bar t \rightarrow \infty}=\int k \wedge
c_{2g-3}(\widetilde{\cal V})\wedge {\cal J}$$
We have not determined this bundle.

\newsec{Speculations and Conjectures}

{}From the discussions in this paper it is clear that the $N=2$ string theory
on $R^2\times T^2$ has a lot of resemblance to the large $N$ description
of 2d Yang-Mills as a string theory \ref\tdqcd{D.J. Gross and W. Taylor,
``Two-Dimensional QCD is a String Theory,'' Nucl. Phys. B400 (1993) 181
\semi D.J. Gross and W. Taylor,
``Twists and Wilson Loops in the String Theory of Two-Dimensional
QCD,'' Nucl. Phys. B403 (1993) 395
\semi S. Cordes, G. Moore, and S. Ramgoolam,`` Large N 2-D
Yang-Mills Theory and Topological String Theory,'' hep-th 9402107\semi
``Lectures on 2-D Yang-Mills Theory, Equivariant Cohomology and Topological
Field Theories,'' Les Houches Summer School proceedings, 1994.}.  Even though
the precise topological computation we ended up with was not
exactly the one appearing in the large N limit of 2d Yang-Mills
 on a torus it is very close to it, in that the primary objects in both cases
is the moduli of
holomorphic maps from Riemann surfaces to a torus.  It is thus
natural to ask if there is any gauge theory which would give
us, as a large $N$ expansion, the $N=2$ string.

In order to narrow down the search we should recall the
natural setting in which the $N=2$ string theory is defined.
First of all the target space dimension of $N=2$ string is four
thus we are looking for a $4d$ gauge theory.  Another fact
motivated from the connection between large N limit of 2d Yang-Mills
theory and string theory is that the latter has no propagating
degrees of freedom and it has only global topological degrees of freedom
(which was the reason for its exact solvability \ref\mig{A. Migdal,
``Recursion Equations in Gauge Theories,'' Sov. Phys. JETP 42
(1975) 413.}).
Another fact is that we are looking for a theory which makes sense
only in two {\it complex} dimensions, as that is the natural
setting for $N=2$ strings.  We only know of one class of gauge
theories which satisfies all these requirements:  It is known
as the {\it holomorphic Yang-Mills theory} in 4d \park\witun.
  It is basically the ordinary Yang-Mills
theory with no matter, but formulated in two complex dimensions and
with the requirement that the field strength
be holomorphic.  This means that, in holomorphic notation,
\eqn\cons{F^{2,0}=F^{0,2}=0}
and the only non-vanishing  component of $F$ is in the $F^{1,1}$ direction.
Of course in general one cannot just set constraints such as \cons\ and
expect to get a consistent field theory.  However, it can
be done in this case \park\witun .
The idea is based on the link established between the 2d Yang-Mills
and Donaldson theory in 2d \ref\wite{E. Witten,``Two Dimensional
Gauge Theories Revisited,'' J. Geom. Phys. 9 (1992) 303.}.
  It
was shown there that ordinary 2d Yang-Mills theory can be
viewed as the deformation of 2d topological Yang-Mills.
Recall that topological Yang-Mills is a twisted version of
$N=2$ Yang-Mills.  The four dimensional analog turns out to
be the natural generalization of this construction:  One starts
from $N=2$ Yang-Mills theory and twists it to obtain
the Donaldson theory \ref\wittend{E. Witten,``Supersymmetric
Yang-Mills Theory on a Four Manifold,'' J. Math. Phys. 35 (1994) 5101.}\
and then perturb it using certain observables of Donaldson
theory and in addition with some topologically trivial deformation.
  In this case one obtains the holomorphic
Yang-Mills theory.  Thus the theory makes sense as a quantum field
theory.

Note that there are no local degrees of freedom in holomorphic
Yang-Mills theory.  To see this we have to go to a Minkowskian version
of the theory.  Consider signature (1,1) in complex notation (which
is the one which also appears for $N=2$ strings).  Let
$$A_i=\int \epsilon_i(p,\bar p) {\rm exp}(i (p\cdot \bar x +
\bar p \cdot x)) $$
$$\bar A_i=\int \bar \epsilon_i(p,\bar p) {\rm exp}(i (p\cdot \bar x +
\bar p \cdot x))$$
where $i=1,2$ denote the holomorphic index.  Then
the linearized equations of motion imply that the support
of $\epsilon$ is on $p\cdot p=0$; moreover the lorentz gauge
condition implies
$$p\cdot \bar \epsilon +\bar p \cdot \epsilon =0$$
This cuts down the real degrees of freedom to 2.  However
we have the constraints \cons\ which imply
$$ \epsilon_{(i} p_{j)}=0$$
where we antisymmetrize in indices,
and this further cuts down the number by two leaving us with
no propagating degrees of freedom as desired\foot{This
same counting also work for one complex dimension, as there
the constraints \cons\ are vacuous.  In complex
dimensions bigger than 2, for generic $p$, the constraints \cons\ will
lead to `negative' number of degrees of freedom.  Thus complex
dimensions 1 and 2 are critical for holomorphic Yang-Mills theory.}.

One other fact which suggests that large $N$ version of gauge theory
in 4d may lead to a string theory is the fact that for finite
$N$ we can turn on `t Hooft magnetic fluxes on the manifold which
live on $H^2(M,{\bf Z_N})$.  As $N\rightarrow \infty$ the choice
of the flux gets related to $H^2(M,U(1))$ which is precisely the
choice of the antisymmetric field $B$ that can be turned on for
string theory that for a fixed $N$ did not have a gauge theory analog.

Putting all this together we feel we have some evidence for the following
conjecture:
\vskip .7cm
{\it The Large} $N$ {\it limit of holomorphic Yang-Mills is
equivalent to} $N=2$ {\it strings}.
\vskip .7cm

How do we check this conjecture?  The natural method should
be first to solve the holomorphic Yang-Mills theory in 4d,
just as Migdall solved the 2d theory.  This should be
an exactly solvable theory as there are no propagating modes,
and steps in solving it have been taken \park .
Another related computation, given the
relation of holomorphic Yang-Mills theory with the
$N=2$ $SU(N)$ Yang-Mills theory, is the large $N$ computations of
Douglas and Shenker
\ref\largen{M. Douglas and S. Shenker, ``Dynamics of SU(N) Supersymmetric
Gauge Theory,'' hep-th/9503163.}.

Perhaps the simplest case to check would be holomorphic Yang-Mills
on $T^4$ (or perhaps $K3$).
Another test, also related to the computation we have done in this
paper is to study the reduction of the holomorphic Yang-Mills
to 2d on a small torus.  Note that in string language
a small torus and a big torus are equivalent by $R\rightarrow 1/R$
so the case we have been considering on $T^2\times R^2$ can be
viewed in this way.  Thus we should get a gauge
theory in 2d which for large $N$ should reproduce
the computations we have done in this paper for all $N$.
Formally this theory should be a deformation of $N=4$ topological
theory in 2d, which at the topological level computes
the Euler characteristic of Hitchin space \ref\hitc{N. Hitchin,
Proc. London Math. Soc. 3 (1987) 55.}.
Note that, being a deformation of a topological theory, it
continues to have no propagating degrees of freedom.
We expect this deformation of 2 dimensional model be exactly
solvable also, just as 2d Yang-Mills is exactly solvable.
This would be very interesting to compute, as the results
of this paper provide an all order prediction for its large $N$ behaviour.

Another point we wish to comment on is whether we can sum up
the perturbation series which we have computed for the
example of $T^2\times R^2$.  Note that in this paper we
computed the amplitudes for each $g$ up to an overall g-dependent
(but modulus independent) constant.  It is tempting to speculate
whether there is a natural choice of the overall constant
which would make the summing up lead to a nice answer.
Even though this may not be a strong test, we have found one
particularly simple choice, which reproduces all $g$ answers,
which agrees with the normalizations we have obtained for $g=1,2$
(up to redefinition of string coupling constant).
Let $\lambda =u^1_Lu^1_R$ and $\bar \lambda= u^2_Lu^2_R$
(we can absorb the definition of string coupling constant into this), then
we have seen in this paper that
$$F^g(\lambda,\bar \lambda)\sim \sum_{(n,m)\not= (0,0)}
|n+m\sigma|^{2g-4}\big(
{\lambda \over n+m\sigma}+{\bar \lambda \over n+m\bar \sigma}\big)^{4g-4}$$
(note in particular that the limit $\tau \rightarrow \infty $ is
proportional to $(\lambda +\bar \lambda)^{4g-4}$, as expected,
only the $m=0$ contributed to the sum).  The natural guess for
summing up all $g$ is thus a geometrical sum
$$\sum_g F^g(\lambda ,\bar \lambda)
=\sum_{(n,m)\not= (0,0)} {1\over |n+m\sigma|^2-(\lambda { \sqrt{n+ m\bar
 \sigma
\over n+m\sigma}}+\bar \lambda {\sqrt{n+m \sigma\over n+m\bar \sigma}})^4}$$
It would be interesting to find out whether this correctly captures
the $g$-dependent constant.

In this paper we have talked about $N=2$ strings which
is equivalent to $N=4$ topological strings, with critical dimension 4.
It would be tempting to connect this with $N=2$ topological strings
which has critical dimension 6 (corresponding to topological
sigma models on Calabi-Yau threefolds).  One idea along this line,
suggested to us by Yau, is to consider the twistor space.
Recall that the twistor space includes a one parameter complex
deformation of the complex structure of the manifold, without
changing the metric.  In particular if $\omega$ denotes the holomorphic
2-form and $k$ the K\"ahler class, then we consider the new holomorphic
2-form $\Omega(t)$ to be
$$\Omega(t) =\omega +t \ k+t^2 {\bar \omega}$$
The total space of the manifold including the parameter
$t$ is a three dimensional complex manifold.  It unfortunately
does not have $c_1=0$.  However, if we turn on an anti-symmetric 2-form
on the three manifold defined by $B=\Omega(t)$ which is a 2-form, we
can modify the condition for conformality from $c_1=0$ because
we now have $H=dB\not= 0$.  In fact we have checked that using
the ideas of the construction of stringy cosmic strings \ref\scs{
B.R. Greene, A. Shapere, C. Vafa and S.-T. Yau,``Stringy Cosmic
Strings and Noncompact Calabi-Yau Manifolds,''
 Nucl. Phys. B337
(1990) 1.}\ that the resulting theory would be a conformal theory.
Thus it may be true that an $N=2$ topological
string on the twistor space, with the $B$-field turned on, is equivalent
to $N=4$ topological string on the 4 manifold.  This we find an extremely
interesting possibility, which deserves further study.

\bigskip

We would like to thank N. Berkovits, R. Dijkgraaf, A. Lesniewski,
G. Moore,
W. Taylor, E. Witten and S.-T. Yau for valuable discussions.  In addition
HO thanks Laurent Baulieu, Marco Picco and the members
of LPTHE, Universit\'e Pierre et Marie Curie where part
of this work was carried out, for their hospitality.
The work of HO was supported
in part by the National Science Foundation under
grant PHY-9501018 and in part by the Director, Office of
Energy Research, Office of High Energy and Nuclear Physics, Division of
High Energy Physics of the U.S. Department of Energy under Contract
DE-AC03-76SF00098.  The work of CV is supported in part
by the NSF grant PHY-92-18167.

\medskip
\appendix{A}{Vanishing of Boundary Terms in the Harmonicity Equation}

In section 3, we have proven that the $N=2$ string amplitude
$F^{g}$ satisfies the harmonicity equation
$$ \epsilon^{ab} u_R^{c} {\partial \over \partial u_L^a}
    {D \over D t^{bc}} F^{g}(u_L, u_R) = 0 $$
provided
\eqn\appnfirst{
\eqalign{ &
\int_{{\cal M}_g}
 \langle (\mu_{A'}, T_L)
 \prod_{A\neq A'} (\mu_A, \widehat{G_L^-}(u_L)) \prod_{A=1}^{3g-3}
(\bar{\mu}_A, \widehat{G_R^-}(u_R)) \times \cr
&~~~~~~~~\times
 \int_\Sigma J_L J_R
 \left[ \int_{\Sigma} \widehat{\tilde{G}^+_L}(u_L)
\widehat{\tilde{G}_R^+}(u_R) \right]^{g-1}
 \int_\Sigma [ \widehat{\tilde{Q}^+_R}(u_R) , \bar{\phi} ]
\rangle \cr}}
and
\eqn\second{
\eqalign{ &
\int_{{\cal M}_g}
 \langle (\mu_{A'}, T_L)(\mu_{A''},J_L^{--})
 \prod_{A\neq A',A''} (\mu_A, \widehat{G_L^-}(u_L)) \prod_{A=1}^{3g-3}
(\bar{\mu}_A, \widehat{G_R^-}(u_R)) \times \cr
&~~~~~~~~\times
 \int_\Sigma J_L J_R
 \left[ \int_{\Sigma} \widehat{\tilde{G}^+_L}(u_L)
\widehat{\tilde{G}_R^+}(u_R) \right]^{g-1}
 \int_\Sigma \{
\widehat{\tilde{Q}_L^+}
(u_L), [ \widehat{\tilde{Q}^+_R}(u_R) , \bar{\phi} ]
\} \rangle \cr}}
vanish. Here we will show that this is indeed the case.
Since the insertion of $(\mu_{A'}, T_L)$ generate a derivative
$\partial/\partial m_{A'}$ on the moduli space ${\cal M}_g$
in the direction of the Beltrami differential $\mu_{A'}$,
we just need to check that
$$ \eqalign{
W_{A'} & =
 \langle
 \prod_{A\neq A'} (\mu_A, \widehat{G_L^-}(u_L)) \prod_{A=1}^{3g-3}
(\bar{\mu}_A, \widehat{G_R^-}(u_R)) \times \cr
&~~~~~~ \times \int_\Sigma J_L J_R
 \left[ \int_{\Sigma} \widehat{\tilde{G}^+_L}(u_L)
\widehat{\tilde{G}_R^+}(u_R) \right]^{g-1}
 \int_\Sigma
[ \widehat{\tilde{Q}^+_R}(u_R) , \bar{\phi} ]
\rangle \cr
V_{A', A''} &=
 \langle (\mu_{A''},J_L^{--})
 \prod_{A\neq A',A''} (\mu_A, \widehat{G_L^-}(u_L)) \prod_{A=1}^{3g-3}
(\bar{\mu}_A, \widehat{G_R^-}(u_R))
 \int_\Sigma J_L J_R \times \cr
&~~~~~~ \times \left[ \int_{\Sigma}
\widehat{\tilde{G}^+_L}(u_L)
\widehat{\tilde{G}_R^+}(u_R) \right]^{g-1}
 \int_\Sigma
\{ \widehat{\tilde{Q}^+_L}(u_L) ,
[ \widehat{\tilde{Q}^+_R}(u_R) , \bar{\phi} ] \}
\rangle \cr} $$
vanish at the boundary of ${\cal M}_g$ whose normal direction
is $\partial_{m_{A'}}$.

As we approach the boundary, the
Riemann surface will degenerate and acquire a node. By conformal
invariance, we can transform the node into a cylinder whose length becomes
infinite at the boundary. In this limit,
$(\bar{\mu}_A, \widehat{G_R^-})$ with $A= A'$ becomes a
contour integral $\oint \widehat{G_R^-}$ around the homology cycle
of the cylinder. Since states propagating along the cylinder are
projected onto zero energy states as we approach the boundary,
they will be annihilated by $\oint \widehat{G_R^-}$ unless
there is another operator on the cylinder which does not
(anti-) commute with $\widehat{G_R^-}$. The only operator
in $V_{A', A''}$ and $W_{A'}$ which does not commute with
$\widehat{G_R^-}$ is $J_R$. However since the commutator
of $J_R$ and $\widehat{G_R^-}$ is proportional to
$\widehat{G_R^-}$ itself, the zero energy states are
still annihilated by $\oint \widehat{G_R^-}$ even if
$J_R$ is inserted on the cylinder. Thus we find that
$V_{A', A''}$ and $W_{A'}$ vanish as we approach the
boundary of the moduli space, and this is what we wanted to show.

\appendix{B}{Genus Two Amplitude on $T^4$}

In this section, we will derive the following
expression of $F^{g}$ at $g=2$ when the target space is $T^4$.
\eqn\expressiongenustwo{\eqalign{ F^{2}(u_L,u_R) =&
\sum_{P_L, P_R} \int_{{\cal M}_2}
\left({\det g \over \det {\rm Im} \Omega} \right)^2
\langle[ \det (\widehat{P}_L+ \widehat{r}_L)
 \det (\widehat{\bar{P}}_R + \widehat{\bar{r}}_R)]^2\rangle  \times \cr
&~~~~~~~~~~~~~~~~~~ \times
\exp[-S(P_L,P_R)] {d^3 \Omega d^3 \bar{\Omega}
\over  [ \det {\rm Im} \Omega]^2}. \cr} }
The notations will be explained in the following.

Let us start with the definition of $F^{2}$
$$  \eqalign{
F^{2} = \int_{{\cal M}_2} & d^3 m d^3 \bar{m}
\langle (\mu_1, \widehat{G_L^-}) (\mu_2, \widehat{G_L^-})
 (\bar{\mu}_1, \widehat{G_R^-}) (\bar{\mu}_2,
   \widehat{G_R^-})  \times \cr
&\times (\mu_3, J_L^{--})( \bar{\mu}_3, J_R^{--})
 [\int \widehat{\tilde{G}_L^+} \widehat{\tilde{G}_R^+} ]^2
\rangle. \cr }
$$
This formula for $F^{2}$ contains {\it two}
$\widehat{G_L^-}$, {\it one} $J_L^{--}$ and
{\it two} $\widehat{\tilde{G}_L^+}$ given by
\eqn\nequalfour{ \eqalign{
& \widehat{G_L^-}=g_{i\bar{j}} \psi_L^{\bar{j}}
\partial \widehat{X}^i \cr
& J_L^{--} = \epsilon_{\bar{i}\bar{j}}
\psi_L^{\bar{i}} \psi_L^{\bar{j}} \cr
& \widehat{\tilde{G}_L^+}= \epsilon_{ij} \psi_L^i \partial
\widehat{X}^j, \cr}}
where
$$ \partial \widehat{X}^i =
   u_L^1 \partial X^i + u_L^2 \epsilon^{ij} g_{j\bar{k}}
          \partial X^{\bar{j}}. $$
Thus there are {\it two} $\psi_L$ and {\it four}
$\bar{\psi}_L$ in $F^{2}$. When the target space is $T^4$
there are $2$ zero modes for $\psi_L$  and $2g$
zero modes for $\bar{\psi}_L$ on genus $g$
since $\psi_L$'s are zero-forms and $\bar{\psi}_L$'s are
one-forms.
Therefore, at genus $2$,
the fermions in $F^{2}$ just absorb their zero modes
and the computation of $F^{2}$  does not involve the Green function.
This will simplify the computation at $g=2$. For $g \geq 3$, we
must deal with the Green function of the fermions.

For the bosonic field $X^i$, we use the decomposition
$$ X^i(z,\bar{z}) = X_0^i(z,\bar{z}) + \phi^i(z,\bar{z}) $$
where $X_0^i$ is the classical part obeying $\partial_z
\bar{\partial}_{\bar{z}} X_0^i = 0$ and $\phi^i$ is the quantum
part which is single-valued on $\Sigma$. The classical
part $X_0^i$ is parametrized by a set of integers
$n^{i}_a, m^{ia}, s^{i}_a, r^{ia}$ ($i,a = 1,2$) as
\eqn\classical{
\eqalign{ X_o^i(z,\bar{z}) =&
 (m^{ia}+ \rho^i_j r^{ja}) {\rm Re} \int^z \omega_a + \cr
 &+ [(n^{i}_a+\rho^i_j s^{j}_a) - (m^{ib}+\rho^i_j r^{jb})
     {\rm Re} \Omega_{ba}] ({\rm Im}\Omega^{-1})^{ac}
{\rm Im} \int^z \omega_c \cr}}
where $\omega_a$ is the holomorphic one-forms on $\Sigma$
and $\Omega_{ab}$ is their period matrix. The matrix $\rho^i_j$
characterize the complex structure of the target space torus
$T^4$ as
$$ T^4 = R^4 / (x^i \sim x^i + n^i + \rho^i_j m^j) ~~~~
(n^i, m^i \in {\bf Z}). $$
The quantum part $\phi^i$ is the free boson whose propagator
is given by
\eqn\quantum{
\eqalign{
  & \langle \partial_z \phi^i \partial_w \phi^{\bar{j}}
\rangle =g^{i \bar{j}}[ -\partial_z \partial_w \log E(z,w) +
    \omega_a(z) \omega_b(w) ({\rm Im} \Omega^{-1})^{ab}] \cr
& \langle \partial_z \phi^i \partial_{\bar{w}}
\phi^{\bar{j}} \rangle =
  - g^{i \bar{j}} \omega_a(z) \bar{\omega}_b(\bar{w})
    ({\rm Im} \Omega^{-1})^{ab} ~~~~(z \neq w),\cr} }
where $E(z,w)$ is the prime form on $\Sigma$. In $F^{2}$,
$\phi^i$ and $\phi^{\bar{i}}$ appear
in the combinations
$$ \eqalign{
  \partial \widehat{\phi}^i & =
   u_L^1 \partial \phi^i +
   u_L^2 \epsilon^{ij} g_{j \bar{k}} \partial \phi^{\bar{k}} \cr
 \bar{\partial} \widehat{\phi}^{\bar{i}} & =
  u_R^1 \bar{\partial} \phi^{\bar{i}} +
  u_R^2 \epsilon^{\bar{i}\bar{j}} g_{\bar{j}k}
    \bar{\partial} \phi^k , \cr} $$
and their Wick contraction rules are
$$ \eqalign{ &
 \langle \partial_z \widehat{\phi}^i \partial_w \widehat{\phi}^j
\rangle = \langle \partial_{\bar{z}}
   \widehat{\phi}^{\bar{i}}
   \partial_{\bar{w}} \widehat{\phi}^{\bar{j}}
  \rangle = 0 \cr
& \langle \partial_z \widehat{\phi}^i
    \bar{\partial}_{\bar{w}} \widehat{\phi}^{\bar{i}}
 \rangle =
- g^{i \bar{j}} (u_L^1 u_R^1 + u_L^2 u_R^2 )
   \omega_a(z) \bar{\omega}_b (\bar{w})
({\rm Im} \Omega^{-1})^{ab}. \cr} $$
Therefore for the purpose of evaluating $F^{2}$, we may write
the bosonic field $X^i$ as
\eqn\boson{
\eqalign{&
\partial \widehat{X}^i = (\widehat{P}_{La}^{i} +
             \widehat{r}_{La}^{i})
({\rm Im} \Omega^{-1})^{ab} \omega_b \cr
& \bar{\partial} \widehat{X}^{\bar{i}} =
(\widehat{P}_{Ra}^{\bar{i}} +
  \widehat{r}_{Ra}^{\bar{i}})({\rm Im} \Omega^{-1})^{ab}
\bar{\omega}_b ,\cr}}
where
\eqn\momentumhat{ \eqalign{&
 \widehat{P}_{La}^{i} = u_L^1 P_{La}^{i} + u_L^2 \epsilon^{ij} g_{j\bar{k}}
    P_{La}^{\bar{k}} \cr
  & \widehat{P}_{Ra}^{\bar{i}} =
  u_R^1 P_{Ra}^{\bar{i}} + u_R^2 \epsilon^{\bar{i}\bar{j}}
g_{\bar{j} k} P_{Ra}^{k} \cr} }
and
\eqn\momenta{ \eqalign{
 P_{La}^{i} & = (n^i_a + \rho^i_j s^j_a) -
                (m^{ib} + \rho^i_j r^{jb}) \bar{\Omega}_{ba} \cr
 P_{La}^{\bar{i}} & = (n^i_a + \bar{\rho}^i_j s^j_a) -
                (m^{ib} + \bar{\rho}^i_j r^{jb}) \bar{\Omega}_{ba} \cr
 P_{Ra}^{i} & =
   (n^i_a + \rho^i_j s^j_a) -
      (m^{ib} + \rho^i_j r^{jb}) \Omega_{ba} \cr
 P_{Ra}^{\bar{i}} & =
   (n^i_a + \bar{\rho}^i_j s^j_a) -
      (m^{ib} + \bar{\rho}^i_j r^{jb}) \Omega_{ba}, \cr} }
and $\widehat{r}_L$ and $\widehat{r}_R$ are quantum variables
obeying the Wick rules
\eqn\reducedwick{
\eqalign{ & \langle \widehat{r}_{La}^{i} \widehat{r}_{Lb}^{j} \rangle
= \langle \widehat{r}_{Ra}^{\bar{i}}
    \widehat{r}_{Rb}^{\bar{j}} \rangle = 0 \cr
& \langle \widehat{r}_{La}^{i} \widehat{r}_{Rb}^{\bar{j}} \rangle
  = - g^{i \bar{j}} (u_L^1 u_R^1 + u_L^2 u_R^2)
({\rm Im} \Omega)_{ab} \cr } }

Now we are ready to evaluate $F^{2}$. Since $\psi_L^i$ and
$\psi_R^{\bar{i}}$  are zero-forms, it is natural
to normalize their zero modes as
$$
\eqalign{& \langle \psi_L^i(z_1) \psi_L^j(z_2) \rangle
   = \epsilon^{ij} \cr
& \langle \psi_R^{\bar{i}}(\bar{z}_1) \psi_R^{\bar{j}}
(\bar{z}_2) \rangle
   = \epsilon^{\bar{i}\bar{j}}. \cr} $$
Thus $\psi_L^i$ and $\psi_R^{\bar{i}}$ may be regarded
as constant on $\Sigma$, and we can perform the surface integral of
$$\eqalign{& \widehat{\tilde{G}_L^+} = \epsilon_{ij} \psi_L^i
 (\widehat{P}_{La}^{j} + \widehat{r}_{La}^{j})
({\rm Im} \Omega)^{ab} \omega_b \cr&
\widehat{\tilde{G}_R^+} = \epsilon_{\bar{i}\bar{j}}
\psi_R^{\bar{i}}
 (\widehat{P}_{Ra}^{\bar{j}} +
   \widehat{r}_{Ra}^{\bar{j}})
({\rm Im} \Omega^{-1})^{ab} \bar{\omega}_b, \cr} $$
as
$$  \int \widehat{\tilde{G}_L^+}
\widehat{\tilde{G}_R^+} =
 \epsilon_{ij} \psi_L^i \epsilon_{\bar{i}\bar{j}} \psi_R^{\bar{i}}
   (\widehat{P}_{La}^{j} + \widehat{r}_{La}^{j})
 (\widehat{P}_{Rb}^{\bar{j}} +
   \widehat{r}_{Rb}^{\bar{j}}) ({\rm Im} \Omega^{-1})^{ab}. $$
The expectation value of these operators then becomes
\eqn\gtilde{
\eqalign{& \langle \int \widehat{\tilde{G}_L^+}
\widehat{\tilde{G}_R^+} \int \widehat{\tilde{G}_L^+}
\widehat{\tilde{G}_R^+}  \rangle = \cr
& = \det( \widehat{P}_L+ \widehat{r}_L)
\det (\widehat{\bar{P}}_R+ \widehat{\bar{r}}_R)
   (\det {\rm Im} \Omega)^{-1} .\cr}}

For $\psi_L^{\bar{i}}$ and $\psi_R^{i}$ zero modes, we can express them
as a linear combination of the holomorphic and anti-holomorphic one-forms
$\omega_a$ ($a=1,2$) as
\eqn\zeromodes{\eqalign{ &  \psi_L^{\bar{i}}(z) =
  \theta_L^{\bar{i}a} \omega_a(z)\cr
    & \psi_R^{i}(\bar{z}) =
  \theta_R^{ia} \bar{\omega}_a(\bar{z})\cr} }
where $\theta_L^{\bar{i}a}$ and
$\theta_R^{ia}$ are Grassmannian variables. We normalize
them as
\eqn\thetanormalization{  \langle \theta_L^{\bar{i}1}
 \theta_L^{\bar{j}1}\theta_L^{\bar{k}2}\theta_L^{\bar{l}2}
\theta_R^{i1}
 \theta_R^{j1} \theta_R^{k2}\theta_R^{l2} \rangle
= \epsilon^{\bar{i}\bar{j}}\epsilon^{\bar{k}\bar{l}}
\epsilon^{ij} \epsilon^{kl}  .}
In evaluating $F^{2}$, we may replace the fermions by their
zero modes \zeromodes\ as
$$ \eqalign{
&\widehat{G_L^-} = g_{i \bar{j}} \theta_L^{\bar{j}a}
   (\widehat{P}_{Lb}^{i}+ \widehat{r}_{Lb}^{i})
({\rm Im} \Omega^{-1})^{bc} \omega_a \omega _c \cr
&J^{--}_L = \epsilon_{\bar{i}\bar{j}}
  \theta_L^{\bar{i}a} \theta_L^{\bar{j}b} \omega_a \omega_b .\cr} $$
Here the following formula becomes useful.
$$ (\mu_A, \omega_a \omega_b) = \int \mu_A \omega_a \omega_b
   = {\partial \Omega_{ab} \over \partial m^A}, $$
where $\partial / \partial m_A$ is a derivative on the moduli
space ${\cal M}_2$ in the direction specified by
the Beltrami-differential $\mu_A$. Thanks to this formula, we can perform
the surface integrals of
$\widehat{G_L^-}$ and $J_L^{--}$ as
$$ \eqalign{&
   (\mu_A, \widehat{G_L^-}) = g_{i\bar{j}}
   \theta_L^{\bar{j}a} (\widehat{P}_{Lb}^{i} + \widehat{r}_{Lb}^{i})
  ({\rm Im} \Omega^{-1})^{bc} {\partial \Omega_{ac}
\over \partial m^A} \cr
 &(\mu_A, J_L^{--}) = \epsilon_{\bar{i}\bar{j}}
  \theta_L^{\bar{i}a} \theta_L^{\bar{j}b} {\partial \Omega_{ab}
 \over \partial m^A}. \cr} $$
By working out simple combinatorics, one finds
$$ \eqalign{
& (\mu_1, \widehat{G_L^-}) (\mu_2, \widehat{G_L^-})(\mu_3, J_L^{--})
   = \cr
& = \theta_L^{\bar{1}1} \theta_L^{\bar{1}2} \theta_L^{\bar{2}1}
    \theta_L^{\bar{2}2} \epsilon_{\bar{1}\bar{2}} \det g
    \det (\widehat{P}_L + \widehat{r}_L ) (\det {\rm Im} \Omega)^{-1}
  {\partial (\Omega_{11}, \Omega_{12}, \Omega_{22} ) \over
    \partial (m_1, m_2, m_3)}. \cr} $$
At genus $2$, so called the Schottky problem is absent,
and we can use the three components of the period matrix
$\Omega_{ab}$ as coordinates on ${\cal M}_2$.
By taking into the normalization of $\theta_L$ and
$\theta_R$ in \thetanormalization, we obtain
\eqn\gandj{ \eqalign{&
\langle (\mu_1, \widehat{G_L^-}) (\mu_2, \widehat{G_L^-})
       (\bar{\mu}_1, \widehat{G_R^-})
(\bar{\mu}_2, \widehat{G_R^-})
 (\mu_3, J_L^{--})(\bar{\mu}_3, J_R^{--})
\rangle d^3 m d^3 \bar{m} = \cr
& = \det g \det(\widehat{P}_L + \widehat{r}_L)
 \det(\widehat{\bar{P}}_R + \widehat{\bar{r}}_R )
{d^3 \Omega d^3 \bar{\Omega} \over [\det {\rm Im} \Omega]^3} \cr} }
where we used $\det g = \epsilon_{12} \epsilon_{\bar{1}\bar{2}}$
to reduce the expression. This relation between $g_{i\bar{j}}$
and $\epsilon_{ij}, \epsilon_{\bar{i}\bar{j}}$
is required in order for the generators
\nequalfour\ to make the $N=4$ superconformal algebra.

Finally, by combining \gtilde\ and \gandj, we derive
$$
\eqalign{ & \langle
  (\mu_1, \widehat{G_L^-})(\mu_2, \widehat{G_L^-})
(\bar{\mu}_1, \widehat{G_R^-})
 (\bar{\mu}_2, \widehat{G_R^-}) \times
\cr &~~\times
(\mu_3, J_L^{--})(\bar{\mu}_3, J_R^{--})
[\int \widehat{\tilde{G}_L^+}\widehat{\tilde{G}_R^+} ]^2
\rangle d^3 m d^3 \bar{m} = \cr
&= \left( {\det g \over \det {\rm Im} \Omega} \right)^2
\left[ \det( \widehat{P}_L+\widehat{r}_L)
   \det(\widehat{\bar{P}}_R+\widehat{\bar{r}}_R) \right]^2
  {d^3 \Omega d^3 \bar{\Omega} \over
[\det {\rm Im} \Omega]^3}. \cr}
$$
One can easily check that this expression is
covariant both on ${\cal M}_2$ and $T^4$.

To complete the evaluation of $F^{2}$, we need to
contract $\widehat{r_L}$ and $\widehat{r_R}$
according to the rule \reducedwick, and multiply
$\exp(-S)$ where $S$ is the classical action for
\classical\ given by
$$ S = \Big( t_{i \bar{j}} P_{La}^i P_{Rb}^{\bar{j}}
   + \bar{t}_{i \bar{j}} P_{La}^{\bar{j}} P_{Rb}^i \Big)
({\rm Im} \Omega)^{ab} ,$$
where
$$\eqalign{& t_{i\bar{j}} = g_{i\bar{j}} +
  i \theta^\alpha k^\alpha_{i \bar{j}} \cr
&\bar{t}_{i\bar{j}} = g_{i\bar{j}} -
  i \theta^\alpha k^\alpha_{i \bar{j}} \cr}$$
and $k^\alpha$ ($\alpha = 1 ,..., h^{1,1}$)
are generators of $H^{1,1}(T^4,{\bf Z})$.
The determinant factors of the bosons $\phi$
and the fermions $\psi_L$ and $\psi_R$ cancel out.
By assembling the ingredients together, we obtain
$$ \int_{{\cal M}_2}
\left( {\det g \over \det {\rm Im} \Omega} \right)^2
\left[ \det( \widehat{P}_L+\widehat{r}_L)
   \det(\widehat{\bar{P}}_R+\widehat{\bar{r}}_R) \right]^2
  {d^3 \Omega d^3 \bar{\Omega} \over
[\det {\rm Im} \Omega]^3} \exp(-S) $$
Finally we sum this over all $n,m,r,s$ parametrizing
$P_L$ and $P_R$ as \momenta.
This way, we have derived the expression \expressiongenustwo\
for $F^{2}$.

\listrefs

\end